# Bend Instabilities and Topological Turbulence in Shear-Aligned Living Liquid Crystal

Hend Baza[1,2], Fei Chen[3], Taras Turiv[1,4], Sergij V. Shiyanovskii[1,4], Oleg D. Lavrentovich[1,2,4*]

[1]Advanced Materials and Liquid Crystal Institute, Kent State University, Kent, OH 44242, USA

[2]Department of Physics, Kent State University, Kent, OH 44242, USA

[3]Department of Mathematical Sciences, Kent State University, Kent, OH 44242, USA

[4]Materials Science Graduate Program, Kent State University, Kent, OH 44242, USA

[*] Corresponding author. Email: olavrent@kent.edu

**Abstract.** Flagellated microswimmers *B. Subtilis* dispersed in a nematic phase of a lyotropic chromonic liquid crystal form a living liquid crystal (LLC). The combination of the passive and active components allows us to analyze how the active component transitions from the shear-imposed alignment into topological turbulence. The lateral extension of the experimental cell is 1000 times larger than the 10-micrometer thickness, to avoid the effect of lateral confinement on the dynamic patterns. The surface anchoring is azimuthally degenerate to avoid permanent anisotropy. After the shear cessation, active forces produced by the microswimmers trigger self-amplifying bend undulations. The amplitude $A$ of undulations grows with time and then saturates, while the wavelength $\lambda$ increases only slightly. Strong bend stresses at the extrema of undulations are released by nucleating ± 1/2 disclination pairs, which multiply to produce topological turbulence. The nucleating pairs and the symmetry axes of the +1/2 disclinations are orientationally ordered. In highly active LLCs, this alignment causes a nonmonotonous time dependence of the disclinations' number, as a fast-moving +1/2 disclination of one pair annihilates a -1/2 disclination of the neighboring pair. The spectra of elastic and kinetic energies exhibit distinct wavevector dependencies caused by the energy cost of the deformed passive nematic background. The study demonstrates how applied shears and a passive viscoelastic background affect the dynamics of active matter. Combined with the previously determined viscoelastic properties of the lyotropic chromonic liquid crystal, the results complete a comprehensive description of the experimentally assessable type of active matter.

## Introduction.

Active nematics are the most studied examples of active matter [1-7]. Their experimental realizations are microtubules activated by molecular motors [8-11], tissue monolayers [12-18], bacterial colonies [19-21], and aqueous suspension of swimming bacteria [22-27]. A uniform active nematic formed by extensile active units, aligned by shear in microfluidic channels [11] or by pressure-driven flows [28], is unstable against the formation of bend undulations in the orientation of neighboring units and chaotic flows called "topological turbulence". These instabilities, predicted theoretically [1, 29-31] are readily observed in aqueous dispersions of bacteria [32-38] and microtubules activated by molecular motors [8, 11, 28].

A different class of active matter has been recently introduced [39], explored [40-53], and reviewed [27, 54, 55] as a living liquid crystal (LLC), also known as a living nematic. In an LLC, an isotropic water solvent is substituted with a (passive) nematic (N), a water-based nontoxic lyotropic chromonic liquid crystal [56-59], thanks to which the solvent acquires viscoelastic properties. In nature, many bacterial types move in viscoelastic fluids [60, 61], some of which, such as mucus [62, 63] and extracellular matrix [64, 65], exhibit orientational order [66] or impart such an order onto the dispersed microswimmers [67-69].

The orientational order and anchoring of the N at the bacterium surface align bacteria with an elongated shape along the local director $\hat{\mathbf{n}}$ of N [39, 70-76], which is apolar, $\hat{\mathbf{n}} = -\hat{\mathbf{n}}$. Although the individual bacteria are polar, their global dispersion in a uniform LLC is not polar: the number of bacteria swimming up and down the director is statistically the same. Apolar LLC behavior is also evidenced by orientational defects in the LLC, disclinations, which are of a semi-integer strength: bacterial orientations change by $\pm\pi$ when one circumnavigates the defect core, thanks to the identity $\hat{\mathbf{n}} = -\hat{\mathbf{n}}$ [39]. Polar flows of bacteria in an LLC, either circular [40, 47], or rectilinear [46], can be created by imposing gradients of $\hat{\mathbf{n}}$ through patterned surface anchoring of the N at bounding plates [77]. These surface anchoring patterns of the passive N command the dynamics of bacteria, producing directional flows [40, 46, 47, 53, 72, 78], and targeted delivery of a micro cargo [46, 73, 74]. However, when the activity exceeds some threshold, the LLC develops dynamic instabilities of orientation [39, 42, 43, 45-47, 52, 53, 79]. Since LLCs combine orientational order, viscoelasticity, and activity in a tunable manner and since they hold promise for applications such as controlled microscale transport, understanding activity-triggered instabilities in them is of fundamental and applied interest.

So far, the LLC has been explored in pre-aligned cells with guiding rails of surface anchoring imposed onto $\hat{\mathbf{n}}$ [39-52]. Such alignment causes permanent anisotropy and makes it difficult to separate the intrinsic properties of LLC from the effect of surface alignment. In this work, to avoid the externally imposed permanent anisotropy, we study the evolution dynamics of an initially shear-aligned LLC to a topologically turbulent LLC in a macroscopically isotropic environment, using flat cells with no pre-imposed in-plane surface anchoring. The initial uniform alignment of the LLC is produced by a temporal unidirectional shear. Once the shear stops, hydrodynamic interactions of bacteria create active flows that produce bend instability and nucleation of disclinations; this behavior is contrasted to the relaxation of the passive N to a uniform state. We measure the undulation wavelength, amplitude, and growth rate of the LLC instabilities as a function of bacterial concentration. The nucleation and evolution of disclinations are quantified by a newly developed Delaunay mesh approach. Experiments enable us to extract density-density, orientation-orientation, and velocity-velocity correlation functions, analyze density fluctuations, and calculate the spectra of elastic energy, kinetic energy, and enstrophy. These multifaceted data present a comprehensive experimental picture of the transition scenario from an oriented LLC to topological turbulence controlled by the balance of active stresses and underlying elasticity in the absence of permanent external anisotropy.

## Methods.

*<u>Living liquid crystal preparation</u>*: We use a *Bacillus Subtilis* strain 1085, which is a rod-shaped bacterium with a head that is 5–7 μm long and ~0.7 μm in diameter. The strain is stored at -80°C. The bacterial colony is grown on Lysogeny broth (Miller composition from Teknova, Inc.) agar plates at 35°C for 12-24 hrs, then transferred to a vial with 10 mL of aqueous Terrific Broth (TB) (Sigma T5574) and grown in a shaking incubator at 35°C for 7-9 hrs. The vials are sealed to increase the resistance of bacteria to oxygen starvation. The bacteria concentration in the growth medium is monitored by measuring optical density. The dispersions are removed from the incubator at the end of the exponential growth stage, at a concentration of $c_0 = 8 \times 10^{14} \text{cells/m}^3$. The bacteria are extracted from the liquid medium by centrifugation.

The N-forming organic component disodium cromoglycate (DSCG) (Spectrum Chemicals, purity >98%) is dissolved in the TB at a concentration 13 wt. %. The solution is a homogeneous nematic at the temperature $(25.0 \pm 0.1)$ °C, at which all experiments are performed. The extracted bacteria are mixed with the DSCG solution to form the LLC.

The chromonic N is not toxic to bacteria [59]. A building unit of the chromonic N is a cylindrical aggregate of ~2 nm in diameter and a length that ranges from 1 to $10^3$ nm [56-58]. The gap between the neighboring aggregates is typically (2-3) nm [80]. For a bacterium such as *B. Subtilis*, with a rodlike body of a length (3–10) μm and a diameter of 1 μm, with attached flagella of a length (5–15) μm, such an environment appears devoid of structure. However, anisotropic surface interactions at the bacterium surface orient the N aggregates along the bacterial axis, so that the swimming direction is collinear with the director $\hat{\mathbf{n}} = -\hat{\mathbf{n}}$ of the passive N [39, 70-76]. In the LLC, the bacteria interact through surface anchoring-mediated elastic forces of the N background and by hydrodynamics, as each bacterium produces two fluid jets directed outwards along its axis [42]. The orientational elasticity and the activity of the LLC could be adjusted independently, by the concentration of chromonic molecules that controls the viscosity and elasticity of the N [81-83] and by the concentration of bacteria, which controls the levels of activity [39].

***Shearing cell.*** A Linkam Optical Shearing System CSS450 with plate-plate geometry is used to shear-align the LLCs. The top and bottom quartz plates are parallel and allow one to observe the sample in a transmission mode. The windows of 2.5 mm in diameter are centered at 6.25 mm from the axis of rotation, Fig. 1. The shearing direction is along the $x$-axis in Fig.1a. The gap distance $h$ between the two plates is 10 μm, which is much smaller than the diameter of the sample, about 7.5 mm. One plate rotates with a controllable angular velocity in the range (0.001-10) s$^{-1}$, which produces shear rates in the range of $0.75 \leq \dot{\gamma} \leq 7500 \, s^{-1}$. Each plate is in thermal contact with a silver heater, which sets the temperature at $(25.0 \pm 0.1)$ °C. In contrast to the previous studies [39, 40, 46, 47], no alignment layers are used; the plates are cleaned with deionized water and acetone before each use and impose degenerate tangential alignment of the N director $\hat{\mathbf{n}}$. In the absence of shear, $\hat{\mathbf{n}}$ assumes any orientation in the plane of cell. To demonstrate this, we subjected the passive N samples of DSCG to shear at a rate of $10 \, s^{-1}$ and then stopped the shear and allowed the material to relax for 15 min. The relaxation resulted in different orientations of the director in the plane of the sample; see Ref. [84] for details.

***Imaging.*** Olympus BX40 and BX50 polarizing optical microscopes equipped with a camera AM Scope MU 2000 (5fps, field of view 730 μm by 486 μm) are used to record videos of the LLCs during and after the shear. We also use polychromatic polarization microscopy (PPM) developed by Shribak [85, 86], equipped with a spectral polarization state generator and an analyzer. An observation window of CSS450 allows one to determine the parameters of interest, such as velocities and orientations of the bacteria, Fig. 1(b,c).

***Analysis.*** All collected videos are processed using a homemade Matlab algorithm to enhance the contrast. Matlab and Mathematica-based codes have been developed to quantify (1) bacterial orientations, (2) the wavelength and amplitude of undulations, (3) positions of bacteria, (4) correlation functions, (5) energy spectra, and (6) density modulations. All codes are described in the Supplementary Information.

To assure statistical reproducibility and robustness of the data, we used standardized protocols and settings (the shear rate, shear chamber geometry, absence of azimuthal anchoring, passive nematic composition, etc.); the only parameter changing was the concentration of microswimmers that varied from $0.5c_0$ to $10c_0$. The LLC behavior was also contrasted to the passive counterpart, with no bacteria. The LLC dynamics was recorded in multiple videos for each concentration. We identified 2000-3000 bacteria in each of the 370-650 frames of each video to extract the system parameters.

## Results and Discussion.

An LLC is a dispersion of motile *B. Subtilis* in an N formed by 13 wt.% solution of DSCG. *B. Subtilis* swims parallel to the local N director by coordinated rotation of about 10-20 helical filaments [39, 74]. A swimming *B. Subtilis* generates a hydrodynamic force dipole[87]: the surrounding fluid is pushed away from the bacterium along the axial direction and pulled towards the bacterium along two perpendicular directions. The development of LLC patterns upon cessation of shear depends on the concentration $c$ of bacteria, varied from $c = 0.5c_0$ to $10c_0$; $c$ controls the activity level. We also explore an inactive N with $c = 0$, as described below.

### 1. Shear-induced patterns
#### 1.1. Shear-induced patterns of passive N.

Under shear, the passive N experiences a cascade of structural transformations, indicative of tumbling behavior [84]. At low shear rates, $\dot{\gamma} \leq 1$ s$^{-1}$, the director $\hat{\mathbf{n}}$ realigns perpendicularly to the shear plane formed by the shear direction (the $x$-axis) and the shear gradient (the $z$-axis), entering the so-called log-rolling regime [84]. As $\dot{\gamma}$ increases to $(1-100)$ s$^{-1}$, disclination loops nucleate [84, 88]. At $\dot{\gamma} > 500$ s$^{-1}$, the texture develops stripes

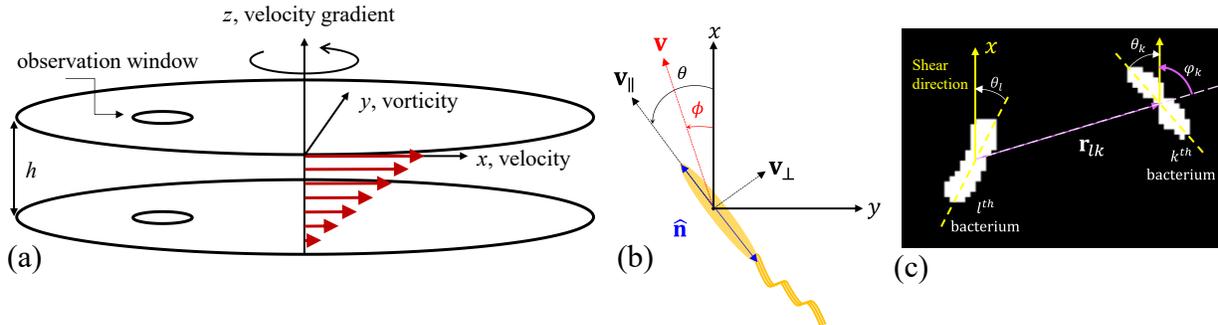

[84], in which $\hat{\mathbf{n}}$ is predominantly along the shear direction, with small periodic clockwise and counterclockwise ≈15° tilts away from the shear plane.

Fig. 1. (a) Scheme of shear; (b) velocities of bacteria; (c) orientational parameters of bacteria.

#### 1.2. Shear-induced patterns of LLC

In the presence of bacteria, $c = (0.5 - 10)c_0$, at a low shear rate $\dot{\gamma} \leq 1\,\text{s}^{-1}$, the LLC director experiences a log-rolling regime, aligning along the $y$-axis, Supplementary Information Fig. S1(a). Shear rates $\dot{\gamma} \sim$ 1-100 s$^{-1}$ produce polydomain textures, Supplementary Information Fig. S1 (b). At $\dot{\gamma} \sim 1000$ s$^{-1}$, the LLC develops stripes, similar to those in the passive N [84], in which the bacteria align along the flow direction, a behavior different from the one reported for isotropic *B. Subtilis* dispersions in water under the shear [89]. In the latter case, shear combined with the chirality of the left-handed flagellum realigns the bacteria counterclockwise, so that the microswimmers acquire a "rheotactic" velocity along the $y$-axis [89]. We did not notice this effect in the LLC, presumably because the shear-aligned director prevents the realignment of bacteria.

## 2. Relaxation upon shear cessation.
### 2.1. Relaxation of passive N upon shear cessation.

Once the shear at $\dot{\gamma} = 1000$ s$^{-1}$ stops at $t = 0$ s, the passive N develops stripes parallel to the shear $x$-direction, which at $t \approx 1$ s are replaced by bands along the $y$-axis, Supplementary Information Fig. S2(a) and Supplementary Video 1. Similar bands have been observed after cessation of flow in liquid crystal polymers [90-92], nematic suspensions of rodlike fd-viruses [93, 94], and cellulose cholesterics [95, 96]. The bands represent bend undulations of the N director; their wavelength $\lambda_0$ increases with time, approximately linearly from ~18 μm at $t = 0.2$ s to 36 μm at $t = 2.4$ s. At $t > 3$ s, the undulations transform into a coarsening polydomain texture, Supplementary Information Fig. S2(a). The relaxation of active LLC upon shear cessation is very different, as described below.

### 2.2. Relaxation of LLC upon shear cessation.

Low concentrations, $c = (0.5 - 2)c_0$, exhibit a peculiar scenario with long-time memory of the shear-aligned state, as the bacteria continue to swim up and down the $x$-axis for a long time, about 3 min after the cessation of flow, Supplementary Information Fig. S3. The elongated bacteria, aligned by shear, anchor the N director $\hat{\mathbf{n}}$, stabilizing the uniform alignment of LLC. The concentrations $c = (0.5 - 2)c_0$ correspond to the bacterial separations $\approx(20-40)$ μm, which are too large for the hydrodynamic force dipoles of neighboring bacteria to overlap and to cause collective behavior and instabilities [47].

High concentrations, $c = 3c_0, 5c_0,$ and $10c_0$ show dynamics illustrated by Supplementary Videos 2, 3, 4, and 5. In these LLC compositions, the separations of bacteria are short, 20 μm or less, permitting hydrodynamic interactions that cause a cascade of instabilities, as illustrated in Fig. 2 for $5c_0$. Upon cessation of shear, at $t = 1 - 2$ s, the bacteria swim up and down the $x$-axis. At $t > 2$ s, the bacterial orientations and the background N director $\hat{\mathbf{n}}$ develop periodic bends, Fig. 2 and Supplementary Information Fig. S2(b). Their period $\lambda$ is in the range $(50 - 110)$ μm, depending on $c$, and is noticeably larger than the wavelength $\lambda_0$ of the short-lived undulations in the inactive N. In what follows, we present details of the activity-induced instabilities in $3c_0$, $5c_0$, and $10c_0$ compositions.

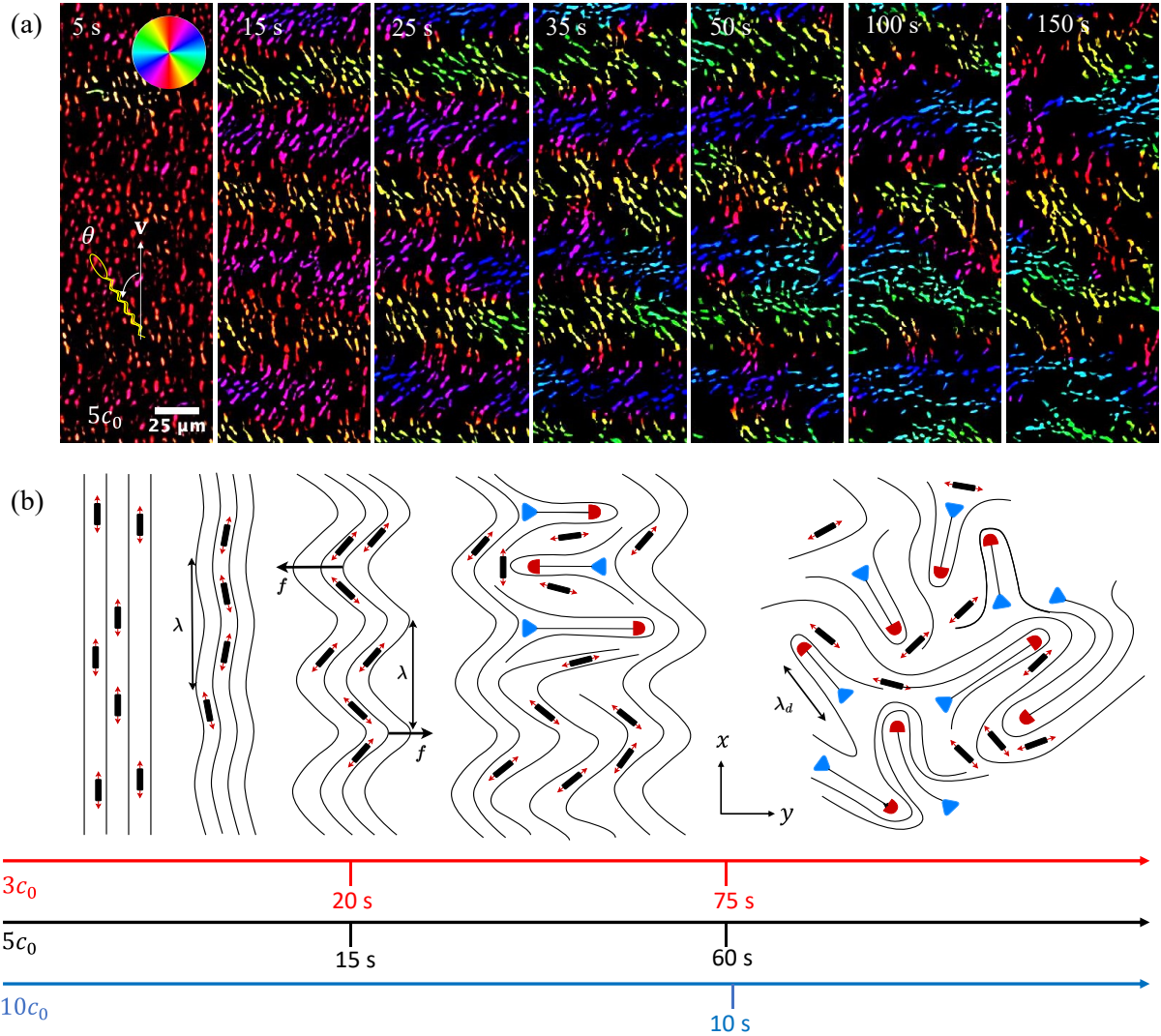

**Fig. 2**. Evolution of the LLC, $c = 5c_0$, after shear cessation: (a) color-coded orientation of bacterial bodies as established by observations under an optical microscope in unpolarized light; (b) scenarios of LLC transition from the uniformly aligned state to bend undulations and topological turbulence for bacterial concentrations $3c_0$, $5c_0$, and $10c_0$, see Supplementary Videos 3, 4 and 5, respectively.

## 3. Bend undulations of LLC upon shear cessation

The orientation of a rodlike bacterium is specified by its axis $\hat{\mathbf{n}}_b = (\cos\theta, \sin\theta)$, where $\theta$ is the angle measured counterclockwise from the shear direction, Fig. 2(a). At the start of relaxation, $\theta \approx 0$, but the range of $\theta$ expands with time, until all reorientations become equally probable in topological turbulence, $t = 150\ s$ in Fig.3. Immediately after the shear cessation, the distribution represents an ellipse elongated along the shear direction. As the bend undulations develop, $t = 50\ s$, the ellipse elongates along the $y$-axis; this reshaping is caused by the tilted orientations of bacteria at the shoulders of the undulation wave. Nucleation of disclination at the ridges of the undulations elongates the ellipse even more, $t = 100\ s$, as the bacterial orientation within the nucleating disclination pairs is along the $y$-axis.

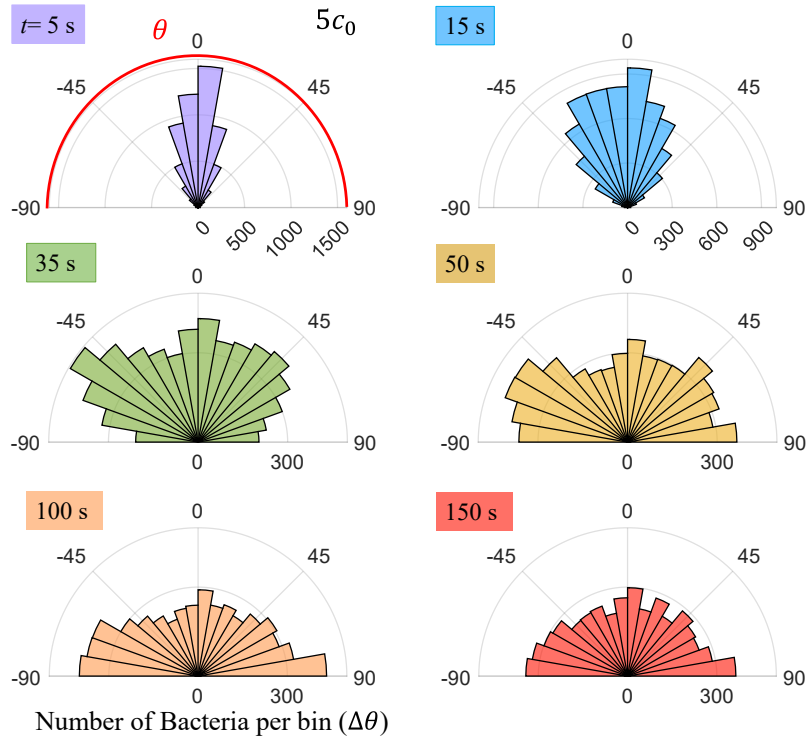

**Fig. 3.** Histograms of bacterial orientation for $5c_0$ dispersion at different times after shear cessation.

Figure 4 illustrates the development of the wavelength $\lambda(t)$, and the amplitude $A(t)$ of undulations for $c = 3c_0, 5c_0$, and $10c_0$. These parameters are determined by fitting the experimentally observed trajectories, which transition from sinusoidal to sawtooth profiles [97], Supplementary Information Fig.S4, and Supplementary Information Section 6. In $5c_0$ and $10c_0$ dispersions, $\lambda$ is practically time-independent; a higher activity brings about a smaller $\lambda$, as $\lambda \approx 75$ μm in $5c_0$ and $\approx 50$ μm in $10c_0$. This behavior is very different from the strong increase in the wavelength $\lambda_0(t)$ of bend in the passive N, see the inset in Fig. 4(a). The behavior of weakly concentrated $3c_0$ LLC is intermediate between strongly active LLC and a passive N, as $\lambda$ slightly increases with time, from $\lambda \approx 80$ μm at the onset of undulations to ~100 μm at later stages.

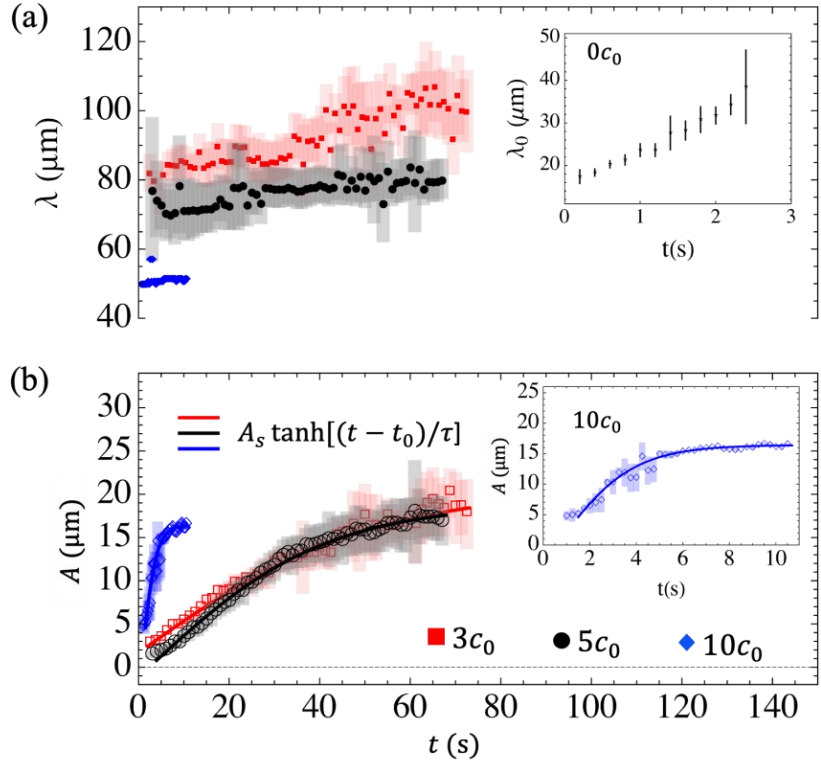

**Fig. 4.** Development of (a) the wavelength $\lambda(t)$ and (b) the amplitude $A(t)$ of undulations for $3c_0$, $5c_0$, and $10c_0$ LLCs. The inset in (a) shows the wavelength $\lambda_0(t)$ of undulations in the passive N after the shear cessation. The inset in (b) shows a fitted $A(t)$ for $10c_0$.

The amplitude $A$, calculated as half the distance between the maximum and minimum of the bending wave, first increases with time and then saturates, Fig. 4(b). The overall growth and saturation could be approximated by the function $A(t) = A_s \tanh\left[\frac{t-t_0}{\tau}\right]$, where $A_s$ is the amplitude at saturation, $\tau$ is the characteristic time at which $A$ starts to saturate, and $t_0$ is the offset time ($t_0 = -4.3$ s for $3c_0$, 2.4 s for $5c_0$, and 0.6 s for $10c_0$), associated with a delay in the undulation onset. The saturation time decreases strongly with the increase of activity/concentration: $\tau \approx (54.2 \pm 2.5)$ s for $3c_0$, $(39.2 \pm 1.1)$ s $5c_0$, and $(3.2 \pm 0.3)$ s for $10c_0$. The amplitudes $A_s$ do not change much: $A_s \approx (20.6 \pm 0.5)$ μm in $3c_0$, $(19.9 \pm 0.3)$ μm in $5c_0$, and $(16.4 \pm 0.3)$ μm in $10c_0$. These values are close to the length scale $\approx (15-20)$ μm at which the hydrodynamic force dipoles of neighboring microswimmers overlap [47].

## 4. Nucleation of defect pairs in LLC.

As the amplitude of undulations increases, while the wavelength remains practically constant, the gradients in the orientational field become stronger, reaching maxima at the extrema of the undulation wave. The increasing stress is relieved by the nucleation of pairs of $\pm 1/2$ disclinations in bacterial orientations. Below, we describe how these defects are identified.

### 4.1. Identification of Defects

To ensure an accurate determination of bacterial orientations, we impose the following criteria on bacterial images: (1) each image is large, with an area above 30 pixels; and (2) the major axis is at least 8 pixels (3.2 μm) and the major axis/minor axis ratio is larger than 1.75. The typical number of bacterial images that satisfy these requirements in one frame ranges from 2600 for $3c_0$ to 3000 for $10c_0$.

Since the N director $\hat{\mathbf{n}}$ is tangential to the bacterial body [42], the long axis $\hat{\mathbf{n}}_b^{(j)} = \left(\cos\theta^{(j)}, \sin\theta^{(j)}\right)$ of a bacterium $j$ making an angle $\theta^{(j)}$ with the shear direction $x$ represents the local director $\hat{\mathbf{n}}(\mathbf{r}^{(j)})$ of the LLC; here $\mathbf{r}^{(j)}$ is the center of mass of a pixelated bacterial image. The identification of disclinations illustrated in Fig. 5(a,b,c) involves the following steps:

(1) The image is fragmented into a set of triangles with vertices at each bacterium location $\mathbf{r}^{(j)}$, Fig. 5(a). We use the Delaunay triangulation algorithm, provided by Mathematica.
(2) A disclination of a strength $m$ implies that circumnavigation around its core results in a director rotation $\Delta\theta = 2\pi m$; counterclockwise circulation and director rotation are considered positive. Using each triangle as a loop, we calculate the director rotation between vertices $k$ and $k'$ as $\theta_{k'k} = \theta_{k'} - \theta_k + l\pi$, where $\theta_k$, k=1,2,3, are the angles of three bacteria at the vertices, and the integer $l$ keeps $\theta_{k'k}$ in the range $(-\pi/2, \pi/2)$. Then, a non-zero value of the total rotation $\Delta\theta = \theta_{21} + \theta_{32} + \theta_{13}$ around the triangle indicates that it contains a disclination of a strength $m = \Delta\theta/2\pi = \pm 1/2$, as illustrated in Fig.5(b). In Fig.5(b), $\theta_1 = -0.20\pi$, $\theta_2 = -0.04\pi$ and $\theta_3 = 0.37\pi$, which result in $\Delta\theta = 0.16\pi + 0.41\pi + 0.43\pi = \pi$ and $m = +1/2$.
(3) The highlighted triangles in Fig. 5(a-c) are of the following types:
   a. Containing an isolated disclination, either $+1/2$ (red triangles) or $-1/2$ (green triangles), surrounded by defect-free triangles, Fig. 5(b). The orientation of an isolated $+1/2$ disclination is defined by the unit vector **p** along the radial director pointing toward the defect core. Since disclinations nucleate in pairs, we introduce a vector **d** that connects a -1/2 disclination to the nearest +1/2 disclination, Fig. 5(b); the pairs are selected by minimizing the total length of distances |**d**| in the frame.

b. Forming bound pairs (BPs), in which a +1/2 triangle shares a common side with a -1/2 triangle; these are marked with magenta and cyan colors, respectively, in Fig.5(c).

c. Forming clusters that contain three or more connected triangles, Fig. 5(a). A cluster of *2n* triangles has an equal number of $\pm 1/2$ defects and can be considered as a connected set of *n* BPs, whereas a cluster of *2n+1* triangles can be considered a connected set of *n* BPs and a triangle with an isolated defect.

The method above allows one to identify the disclinations and analyze their evolution, as discussed below.

### 4.2. *Evolution of Defects.*

Figure 5(a) and Supplementary Information Fig. S5 present selected frames of defects' evolution. Each frame exhibits the numbers $n_{BP}$, $n_+$, and $n_-$ of BPs and isolated $\pm 1/2$ disclinations, respectively. A small random imbalance of $n_+$, and $n_-$ results from the motion of disclinations across the frame borders.

The evolution of disclinations is characterized by the number densities of the +1/2 disclinations $\rho_+ = n_+/\mathcal{A}$ and of the BP pairs $\rho_{BP} = n_{BP}/\mathcal{A}$, Fig.5(e), and by the orientational order of isolated $\pm 1/2$ disclinations, Fig.5(f,g), defined through the averages of the unit vectors **p** and $\hat{\mathbf{d}} = \mathbf{d}/|\mathbf{d}|$, Fig.5(b). Here $\mathcal{A}$ is the frame area. We start the analysis with the $3c_0$ and $5c_0$ LLCs that exhibit similar evolution.

$3c_0$ **and** $5c_0$ **LLC dispersions.** At the early stages after shearing cessation, $t \leq 10$ s, only a few BPs form; there are no isolated $\pm 1/2$ disclinations, Fig. 5(e) and Supplementary Information Figs. S5(a,d). At $t > 10$ s, $\rho_{BP}$ increases, and at $t > 15$ s, repulsion of $+1/2$ and $-1/2$ disclinations of BPs produces isolated disclinations, Figs. 5(a,b) and Supplementary Figs. S5(b,e). The densities $\rho_+$ and $\rho_{BP}$ in $3c_0$ and $5c_0$ dispersions grow with time, Fig. 5(e). In the well-developed topological turbulence, $t > 60$ s, $\rho_{BP}$ continues to increase, while $\rho_+$ saturates in the $3c_0$ and exhibits a small decay in $5c_0$ system, Fig. 5(e). This decay in the density of isolated disclinations can be attributed to the well-developed active flows that bring +1/2 and -1/2 disclinations together and stimulate their annihilation.

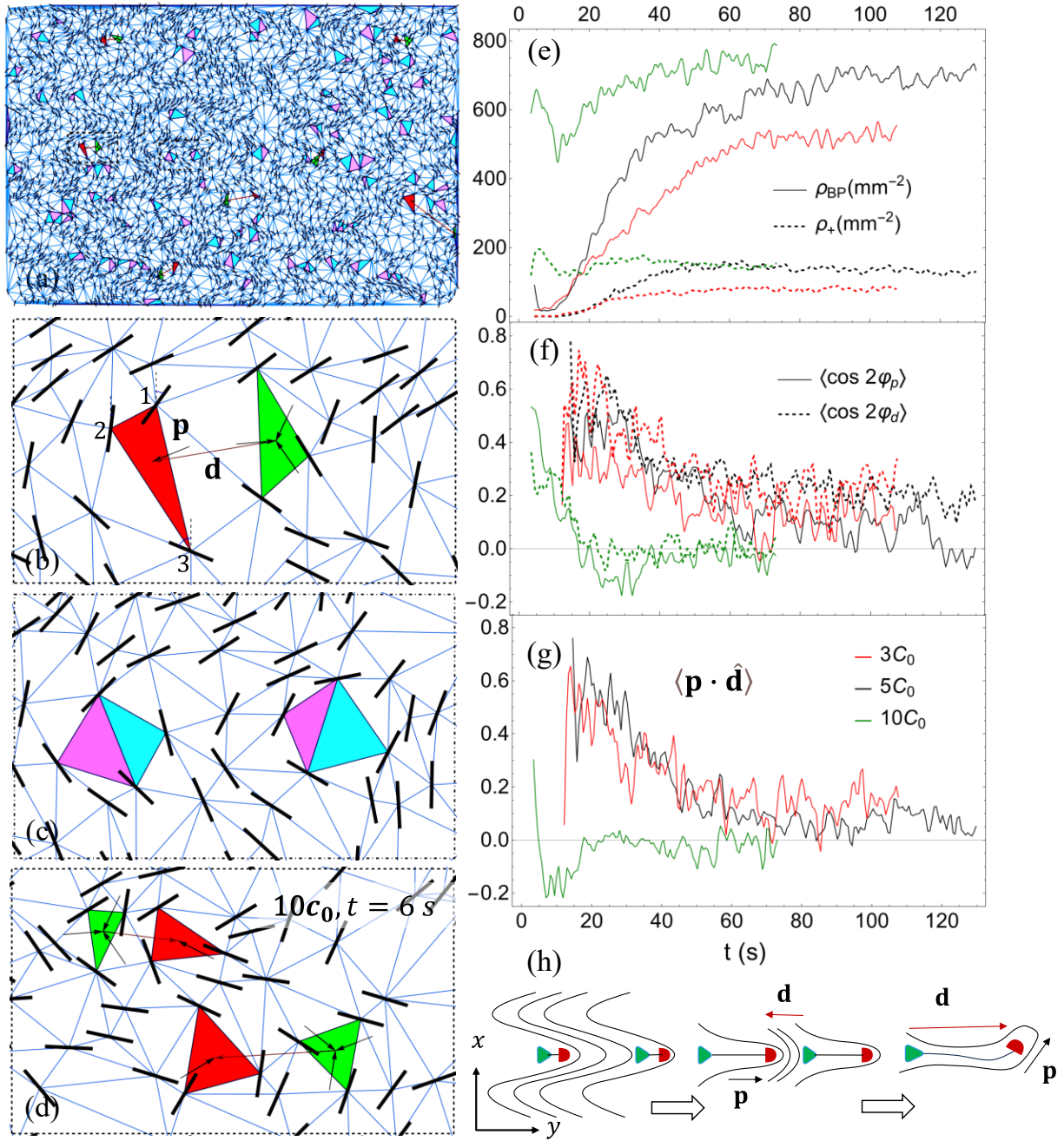

**Fig. 5.** Identification and evolution of disclinations in an LLC: (a) a frame at the onset of topological turbulence contains $n_+$ red triangles (containing $+1/2$ disclinations), $n_-$ green ones (with $-1/2$ disclinations), and $n_{BP}$ BPs composed of magenta ($+1/2$) and cyan ($-1/2$) triangles; (b) fragment of frame (a) where the orientation of $+1/2$ disclination is defined by the unit vector $\mathbf{p}$ along the radial director toward the defect core and the vector $\mathbf{d}$ connects the $-1/2$ to the nearest $+1/2$ disclinations; (c) fragment of frame (a) with BPs; (d) fragment of Fig.S5(g) where the pair selection connects the disclinations nucleated in different events. The evolution of disclinations for $3c_0$ (red), $5c_0$ (black), and $10c_0$ (green) dispersions: (e) number densities of the +1/2 disclinations $\rho_+$ and of the BP pairs $\rho_{BP}$; (f) the orientational order of +1/2 disclinations $\langle\cos2\varphi_p\rangle$ and of unit vectors $\hat{\mathbf{d}} = \mathbf{d}/|\mathbf{d}|$ $\langle\cos2\varphi_d\rangle$; (g) average of the mutual orientation of the paired disclinations $\langle\mathbf{p}\cdot\hat{\mathbf{d}}\rangle$; **(h)** a scheme showing how +1/2 and -1/2 disclinations from two different pairs annihilate as a result of self-propulsion of the +1/2 defect and how the $\mathbf{p}$ vector of +1/2 disclination loses its initial orientation along the $y$-axis through rotational diffusion.

At the onset of topological turbulence, Figs. 5(a,b), the isolated +1/2 disclinations and the defect pairs align along the $y$-axis, as $\langle\cos2\varphi_p\rangle \approx 0.4$ and $\langle\cos2\varphi_d\rangle \approx 0.6$, Fig.5(f). The background N director in the nucleating pair is parallel to the line connecting the two cores: $\langle\mathbf{p}\cdot\hat{\mathbf{d}}\rangle \approx 0.6$, Fig. 5(g). This director is perpendicular to the shear direction, which implies that the nucleating defects randomize the alignment introduced by the shear. As time progresses, the orientational order weakens: $\langle\mathbf{p}\cdot\hat{\mathbf{d}}\rangle$ and $\langle\cos2\varphi_p\rangle$ decrease to zero, while $\langle\cos2\varphi_d\rangle$ remains finite, $\approx 0.2$, even at the latest times, $t$=124 s, Fig. 5(f). The observations suggest that the orientational diffusion of $\mathbf{p}$ is much faster than the translational diffusion of +1/2 cores along the $x$-axis, Fig. 5(h).

**$10c_0$ LLC dispersions.** The evolution of the highly active $10c_0$ LLC differs substantially from the $3c_0$ and $5c_0$ LLCs by a nonmonotonic time dependence of $\rho_{BP}$ and $\rho_+$, Fig. 5(e). At $t = 3$ s, multiple defect pairs have already nucleated, preserving orientation of +1/2 disclinations along the $y$-axis, as $\langle\cos2\varphi_p\rangle \approx 0.5$ is larger than $\langle\cos2\varphi_d\rangle \approx 0.3$, and $\langle\mathbf{p}\cdot\mathbf{d}\rangle \approx 0.3$, Figs. 5(f,g). Both $\rho_{BP}$ and $\rho_+$ peak at $t \approx 5$ s, then undergo a dramatic decrease, reaching minima at $t \approx 11$ s, then grow at $t > 15$ s and eventually saturate in the well-developed topological turbulence, Fig. 5(e). The evolution of $\rho_{BP}$ and $\rho_+$ correlates with the behavior of $\langle\mathbf{p}\cdot\hat{\mathbf{d}}\rangle$ in Fig. 5(g). The nonmonotonous dynamics of disclinations is illustrated in Fig. 5(h) and explained below.

In the strongly active LLC, the nucleating defect pairs align along the $y$-axis. The +1/2 part of the defect pair is mobile; it propels along $\mathbf{p}$ [98]. The -1/2 defects are less mobile because of their symmetry. In the $10c_0$ dispersion, both the concentration of nucleating pairs and the speed of +1/2 cores, proportional to the activity, are high. The fast-moving +1/2 defect runs into the -1/2 defect of a nearby pair and annihilates with it, Fig. 5(h). This "alignment and activity-triggered annihilation" decreases the defect density within the time interval (6-20) s, Fig.5(e). Annihilation implies that the separation between defects temporarily increases; fast orientational diffusion of +1/2 disclinations, Fig. 5(h), hinders further annihilations. The annihilation of +1/2 disclinations aligned along the $y$-axis happens within the $t = (20 - 40)$ s interval: $\langle\cos2\varphi_p\rangle \approx -0.15$ is negative, which means that defects that did not annihilate orient predominantly along the shear $x$-direction rather than along the $y$-axis. Misalignment of defects and a lower probability of annihilation, Fig. 5(f), increases their numbers, Fig.5(e). This alignment and activity-triggered annihilation is absent in $3c_0$ and $5c_0$ since the nucleating defects are separated by longer distances.

The quantitative analysis of the LLC instabilities presented above establishes a connection between the number densities of disclinations and their orientational order and uncovers a new effect of the alignment and activity-triggered annihilation in highly active dispersions, Figs. 5(e-g).

### 5. Velocity analysis

The velocity $\mathbf{v}_n^{(j)}$ of an $j^{\text{th}}$ bacterium in a frame $n$ is calculated as $\mathbf{v}_n^{(j)} = \left[\left(\mathbf{r}_n^{(j)} - \mathbf{r}_{n-1}^{(j)}\right) + \left(\mathbf{r}_{n+1}^{(j)} - \mathbf{r}_n^{(j)}\right)\right]/2\Delta t$, where $\mathbf{r}_n^{(j)}$ is the bacterium's location in the $n^{\text{th}}$ frame. A bacterium swims along $\hat{\mathbf{n}}_b^{(j)} = \hat{\mathbf{n}} = (\cos\theta, \sin\theta)$, where $\theta$ is the angle between the bacterium and the $x$-axis. In the absence of other bacteria, $\mathbf{v}^{(j)}$ is collinear with $\hat{\mathbf{n}}^{(j)}$. However, the velocities $\mathbf{v}^{(j)}$ are often observed at a substantial angle to $\hat{\mathbf{n}}^{(j)}$. The perpendicular $\mathbf{v}_\perp^{(j)}$ component of $\mathbf{v}^{(j)}$ is attributed to the active flows produced by bacterial hydrodynamic interactions; these flows advect a bacterium along a direction different from $\hat{\mathbf{n}}^{(j)}$. We thus represent $\mathbf{v}^{(j)}$ as a sum of a component parallel to $\hat{\mathbf{n}}^{(j)}$, $\mathbf{v}_\parallel^{(j)} = \left(\hat{\mathbf{n}}^{(j)} \cdot \mathbf{v}^{(j)}\right)\hat{\mathbf{n}}^{(j)}$, and $\mathbf{v}_\perp^{(j)} = \mathbf{v}^{(j)} - \mathbf{v}_\parallel^{(j)}$, Supplementary Information Section 9 and Fig. S6.

In a sheared dispersion, $\theta = 0$ and $\mathbf{v}_\perp^{(j)} = 0$, Fig. 3(a). Once the shear stops, a fluctuating bend in the orientation of neighboring bacteria is enhanced by the mechanism proposed by Simha and Ramaswamy [29, 99] and illustrated in Fig. 3(b). Each bacterium is a pusher that produces two antiparallel fluid jets along $\hat{\mathbf{n}}^{(j)}$. At short separations of bacteria, ≤ (15-20) μm, these jets overlap [47]. For bacteria on the two shoulders of the bend wave, the overlaps produce a flow orthogonal to the $x$-axis. This flow amplifies the bend amplitude, while leaving the wavelength $\lambda$ practically intact; the latter is defined by the balance of active forces which favor a shorter $\lambda$ and the N elasticity which favors a longer $\lambda$ [39].

The individual velocity $\mathbf{v}^{(j)} = \mathbf{v}_s^{(j)} + \bar{\mathbf{v}}_a$ is the sum of the swimming velocity $\mathbf{v}_s^{(j)} = v_s^{(j)}\hat{\mathbf{n}}^{(j)}$ and the active flow $\bar{\mathbf{v}}_a$; $\bar{\mathbf{v}}_a$ can be determined by fitting the individual velocities if $\bar{\mathbf{v}}_a$ is assumed to be the same for neighboring bacteria:

$$\mathbf{v}_\perp^{(j)} = \mathbf{v}_s^{(j)} - \left(\hat{\mathbf{n}}^{(j)} \cdot \mathbf{v}_s^{(j)}\right)\hat{\mathbf{n}}^{(j)} = \bar{\mathbf{v}}_a - \left(\hat{\mathbf{n}}^{(j)} \cdot \bar{\mathbf{v}}_a\right)\hat{\mathbf{n}}^{(j)}. \tag{1}$$

The evolution of color-coded averaged $y$-component of the active flow velocity $\bar{v}_{ay}$, obtained by fitting $\mathbf{v}_\perp^{(j)}$ with Eq.(1) within a disc of a radius 20 μm, Fig. 6(a), shows a strong correlation with patterns of orientation in undulations and turbulence, Fig.3. The active flows are not simple vector summations of the individual swimming velocities, Fig.6(b). Since bacteria are force-free swimmers, their propulsion involves two forces of the same magnitude but opposite orientation. The propulsion force $\mathbf{F}_{prop}^{(j)}$ is parallel to $\mathbf{v}_\parallel^{(j)}$ while the drag force $\mathbf{F}_{drag}^{(j)}$ is antiparallel to $\mathbf{v}_\parallel^{(j)}$. The active flow $\bar{\mathbf{v}}_a$ could result from the addition of $\mathbf{F}_{prop}^{(i)}$ and $\mathbf{F}_{prop}^{(j)}$, or $\mathbf{F}_{drag}^{(i)}$ and $\mathbf{F}_{prop}^{(j)}$, or $\mathbf{F}_{drag}^{(i)}$ and $\mathbf{F}_{drag}^{(j)}$. For example, scenarios 3 and 4 in Fig. 6(b) show two large vectors $\mathbf{v}_\parallel^{(i)}$ and $\mathbf{v}_\parallel^{(j)}$ directed to the right while the active flow $\bar{\mathbf{v}}_a$ is directed to the left.

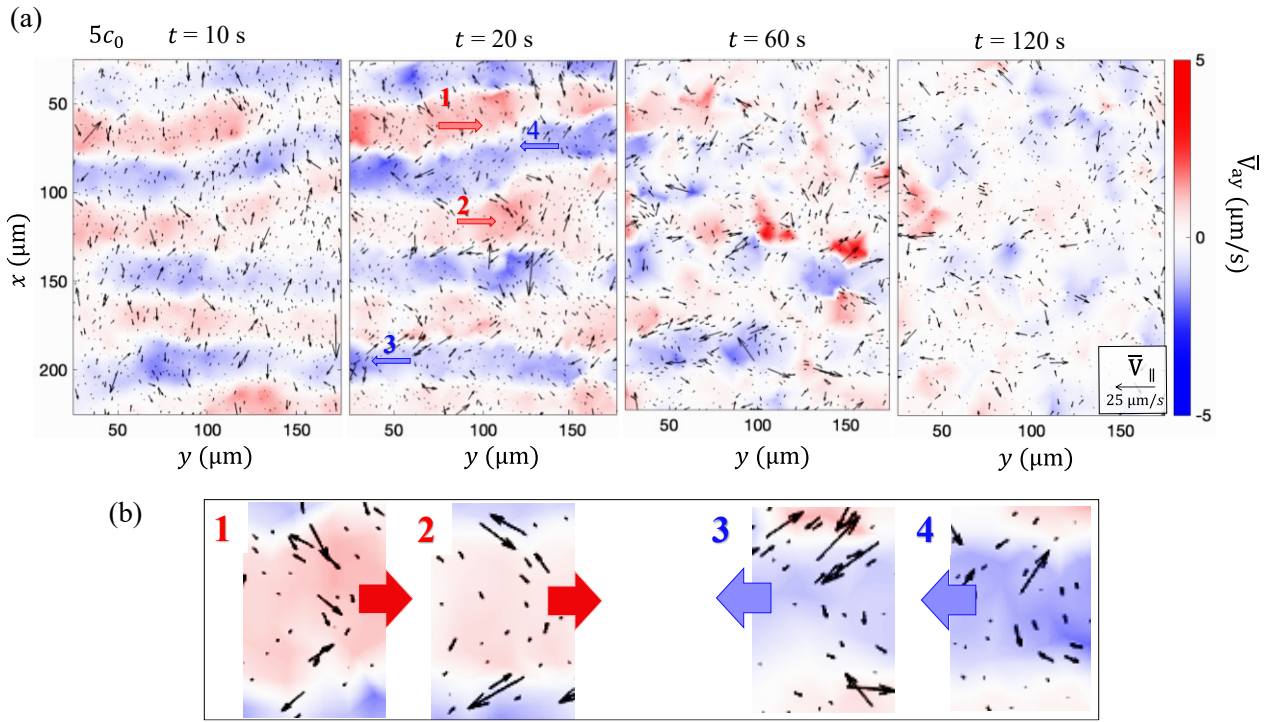

**Fig. 6.** (a) The color-coded averaged horizontal component $\bar{v}_{ay}$ of the orthogonal velocity overlayed with the parallel velocities of bacteria $\mathbf{v}_\parallel^{(j)}$ shown as black arrows at $t$ = 10, 20, 60, and 120 s in $5c_0$; (b) a few enlarged scenarios (taken at $t$ = 20 s) demonstrating that the polarity of the active flow $\bar{\mathbf{v}}_a$ is not a result of a simple vector summation of individual velocities $\mathbf{v}_\parallel^{(j)}$ of bacteria. Scenarios 1 and 2 involve an active flow to the right, while 3 and 4 involve an active flow to the left.

## 6. Analysis of correlations

In this subsection, we analyze the density-density, orientation-orientation, and velocity-velocity correlation functions.

***Pair correlation function***. We start with the pair distribution function $P_2(r, \varphi, \theta_l, \theta_k)$ of finding two bacteria $k$ and $l$, oriented along the directions $\theta_l$ and $\theta_k$, and separated by a distance $\mathbf{r} = r(\cos\varphi, \sin\varphi)$; here, the angles $\varphi$, $\theta_k$ and $\theta_l$ are calculated counterclockwise from the shear direction $x$, Fig. 1(c). Using the asymptotic behavior $P_2(r \to \infty, \varphi, \theta_l, \theta_k) \to \rho P_1(\theta_l)P_1(\theta_k)$, where $\rho$ is the density of bacteria per unit area and $P_1(\theta)$ is the single bacterium's orientational function that obeys the condition $\int_0^{2\pi} P_1(\theta)d\theta = 1$, we introduce the dimensionless correlation function $C(r, \varphi, \theta_l, \theta_k) = \rho^{-1} P_2(r, \varphi, \theta_l, \theta_k) - P_1(\theta_l)P_1(\theta_k)$ and represent it as a Fourier series

$$C(r, \varphi, \theta_l, \theta_k) = \frac{1}{(2\pi)^3} \sum_{n_l, n_k, n=-\infty}^{\infty} C_{n_l, n_k}^{(n)}(r) \exp[i(n_l \theta_l + n_k \theta_k + n\varphi)], \quad (2)$$

where the Fourier amplitudes are integral characteristics of $C(r, \varphi, \theta_l, \theta_k)$:

$$C_{n_l, n_k}^{(n)}(r) = \iiint_0^{2\pi} C(r, \varphi, \theta_l, \theta_k) \exp[-i(n_l \theta_l + n_k \theta_k + n\varphi)] \, d\theta_l d\theta_k d\varphi . \quad (3)$$

The indices $n_{l,k}$ define the type of bacterium's characteristics that the correlation function quantifies. A bacterium is polar, with polarity defined by the swimming direction, Fig. 1(c). For $n_k = 0$, the $k^{th}$ bacterium is a point-like object; the corresponding correlation function quantifies the density correlations. The terms with $n_k = \pm 1$ describe polar correlations caused by the bacterium's polarity. The terms with $n_k = \pm 2$ describe the quadrupolar correlations induced by the elongated bacterial body; the body could be represented as an ellipse. The index $n$ defines the azimuthal dependence of the corresponding contribution to the correlation function. The Fourier amplitudes $C_{n_l, n_k}^{(0)}(r)$ with $n=0$ describe the azimuthally averaged correlations; $C_{n_l, n_k}^{(1)}(r)$ with $n=1$ are not calculated since we do not distinguish between the head and the tail of a bacterium; $C_{n_l, n_k}^{(2)}(r)$ with $n=2$ describe quadrupolar correlations; these are different for two particles placed in a director field side-by-side and for two particles placed head-to-tail. The amplitudes $C_{n_l, n_k}^{(n)}(r)$ exhibit the following properties: (a) $C_{n_l, n_k}^{(n)}(r) = \left(C_{-n_l, -n_k}^{(-n)}(r)\right)^*$ since $C(r, \varphi, \theta_l, \theta_k)$ is real, (b) since the system is globally apolar, $C_{n_l, n_k}^{(n)}(r) = 0$ when $n_l + n_k + n$ is odd, (c) since the system shows a global mirror symmetry, $x \leftrightarrow -x$, then $C_{n_l, n_k}^{(n)}(r) = C_{-n_l, -n_k}^{(-n)}(r)$ is real when $n_l + n_k + n$ is even. We neglect the terms with any of $|n_l|, |n_k|$, and $|n|$ being larger than 2, as

## 6.1. Density-density and orientation-orientation correlations.

The density-density correlation function $C_{dd}(r,\varphi)$ corresponds to $n_l = n_k = 0$, and $n = 0; 2$. It can be represented as the sum of isotropic and anisotropic terms:

$$C_{dd}(r,\varphi) = C_{00}^{(0)}(r) + C_{00}^{(2)}(r)\cos 2\varphi, \tag{4}$$

where the isotropic part $C_{00}^{(0)}(r) = \frac{1}{2\pi}\int_0^{2\pi} C_{dd}(r,\varphi)\, d\varphi$ describes the radial dependence of the azimuthally averaged density-density correlation function and the azimuthal amplitude

$$C_{00}^{(2)}(r) = \frac{1}{\pi}\int_0^{2\pi} C_{dd}(r,\varphi)\cos(2\varphi)\, d\varphi \tag{5}$$

exhibits the azimuthal dependence of $C_{dd}(r,\varphi)$, namely, $C_{00}^{(2)}(r) > 0$ when $C_{dd}(r,0) > C_{dd}(r,\pi/2)$ and vice versa.

The isotropic part $C_{00}^{(0)}(r)$ of the density-density correlation in $5c_0$ LLC is negative at short distances $r < 5$ μm and shows a pronounced maximum at $r \approx 7$ μm at all times, Fig.7(a). The negative values of $C_{00}^{(0)}(r)$ are associated with the "steric" repulsions of the bacteria, which are caused not only by the direct contacts but also by the repulsions through hydrodynamic and orientational elastic forces of the N background. The length 7 μm is a typical length of the bacterium's head. We thus associate the maximum of $C_{00}^{(0)}(r)$ with the tendency of bacteria to form chains[42]. A swimming bacterium leaves a track in the N background that is the least-resistance direction for the bacteria that follow. The chaining effect is evidenced by the behavior of the anisotropic parts of the density-density correlations, Fig.7(b). The corresponding function $C_{00}^{(2)}(r)$ is negative for $r < 5$ μm right after the shear cessation when the bacteria are still oriented along the $x$-axis. At $t =$(15-30) s, $C_{00}^{(2)}(r)$ shows a small positive maximum at $r \approx 10$ μm. At $t > 30$ s, when the defects nucleate and realign the N director by 90°, $C_{00}^{(2)}(r)$ exhibits a negative minimum at 10 μm $< r <$ 20 μm, clearly seen in Fig.7(b). The azimuthally dependent density-density correlation $C_{00}^{(2)}(r)$ shows an interesting far-field behavior, with two positive maxima at $r = 40$ μm and 80 μm, and two minima at 60 μm and 100 μm. These extrema of $C_{00}^{(2)}(r)$ correlate clearly with the wavelength $\lambda \approx 80$ μm of the bend undulations in Fig.4(a).

The quadrupole-quadrupole correlation function is defined as

$$C_{qq}(r,\varphi) = \rho^{-1}\iint_0^{2\pi} P_2(r,\varphi,\theta_l,\theta_k)\cos(2\theta_l - 2\theta_k)\, d\theta_l d\theta_k - S^2, \tag{6}$$

where $S = \int_0^{2\pi} P_1(\theta_l)\cos(2\theta_l)d\theta_l$ is the scalar order parameter of the bacterial array calculated with respect to the shear $x$-axis. It could be written in a form similar to the density-density correlation function, Eqs. (4,5)

$$C_{qq}(r,\varphi) = C_{22}^{(0)}(r) + C_{22}^{(2)}(r)\cos 2\varphi, \qquad (7)$$

where $C_{22}^{(0)}(r)$ and $C_{22}^{(2)}(r)$ are, respectively, the azimuthal average and the azimuthal amplitude of the quadrupole-quadrupole correlation function $C_{qq}(r,\varphi)$; $C_{22}^{(2)}(r) > 0$ when $C_{qq}(r,0) > C_{qq}(r,\pi/2)$ and vice versa.

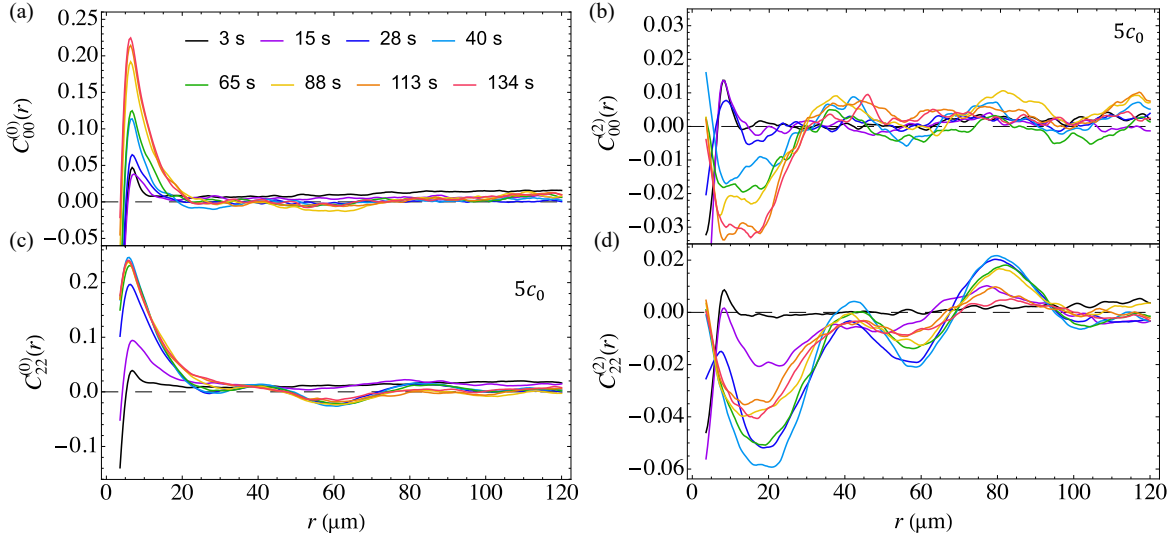

**Fig. 7.** The spatial density-density and the orientation-orientation correlation functions as a function of $r$ at different times for $5c_0$: (a) the azimuthal average $C_{dd}^{(0)}(r)$ and (b) the azimuthal amplitude $C_{dd}^{(2)}(r)$ of the density-density correlation function, Eq.(4); (c) the azimuthal average $C_{qq}^{(0)}(r)$ and (d) the azimuthal amplitude $C_{qq}^{(2)}(r)$ of the quadrupole-quadrupole correlation function, Eq. (7).

For small separations $r < 30$ μm, $C_{22}^{(0)}(r)$, Fig. 7(c), and $C_{22}^{(2)}(r)$, Fig. 7(d), follow $C_{00}^{(0)}(r)$ and $C_{00}^{(2)}(r)$, respectively, Fig. 7(a,b), since the bacteria are parallel to each other. The far-field correlations within the time interval $15\,\text{s} < t < 60\,\text{s}$ are nonmonotonous with maxima at $r \approx 40$ μm and $\approx 80$ μm, Fig. 7(c), which correlates with the undulation wave in bacterial orientations. The radial modulations of $C_{22}^{(k)}(r)$ in Fig. 7(c,d) are stronger than the similar modulations of $C_{00}^{(k)}(r)$ in Fig. 7(a,b), which means that the modulations are indeed caused by the undulations with the period $\lambda \approx 80$ μm. At later stages, $t > 60$ s, in topological turbulence with vortices, (i) the radial modulations of $C_{22}^{(0)}(r)$ indicate that the average distance between the vortices from the same pair corresponds to the periodicity of initial undulations, Fig. 7(c); (ii) the variation of anisotropic term preserves the structure typical for earlier stages of the undulations, Fig. 7(d), and corroborates the long-term memory of the pair orientation $\langle \cos 2\varphi_d \rangle$, Fig. 7(f).

### *6.2.    Velocity-Velocity Correlation.*

We calculate the velocity-velocity correlation function $C_{vv}(R) = \langle \bar{v}_{\perp y}(r)\bar{v}_{\perp y}(r+R)\rangle$ for the quantity $\bar{v}_{\perp y}$, which is of the $y$-component of the perpendicular velocity $\bar{v}_{\perp}$. We represent $C_{vv}(R)$ as two separate parts: one depends on the separation projected onto the shear direction, $C_{vv}(R_x) = \langle \bar{v}_{\perp y}(x)\bar{v}_{\perp y}(x+R_x)\rangle$, and the second depends on the separation along the vorticity direction, $C_{vv}(R_y) = \langle \bar{v}_{\perp y}(y)\bar{v}_{\perp y}(y+R_y)\rangle$. Figure 8(a) shows that during the bend instability, $t < 60$ s, $C_{vv}(R_x)$ reaches a negative minimum at $R_x \approx 40$ µm, caused by an anticorrelation between the flow velocities at the neighboring extrema of the undulation wave, and a maximum at $R_x \approx 80$ µm, which corresponds to the full wavelength $\lambda \approx 80$ µm of undulations, Fig.8(a). As the LLC transitions to turbulence at $t > 60$ s, $C_{vv}(R_x)$ and $C_{vv}(R_y)$ behave similarly, as expected for a system that lost the memory of its original orientation and undulation waves.

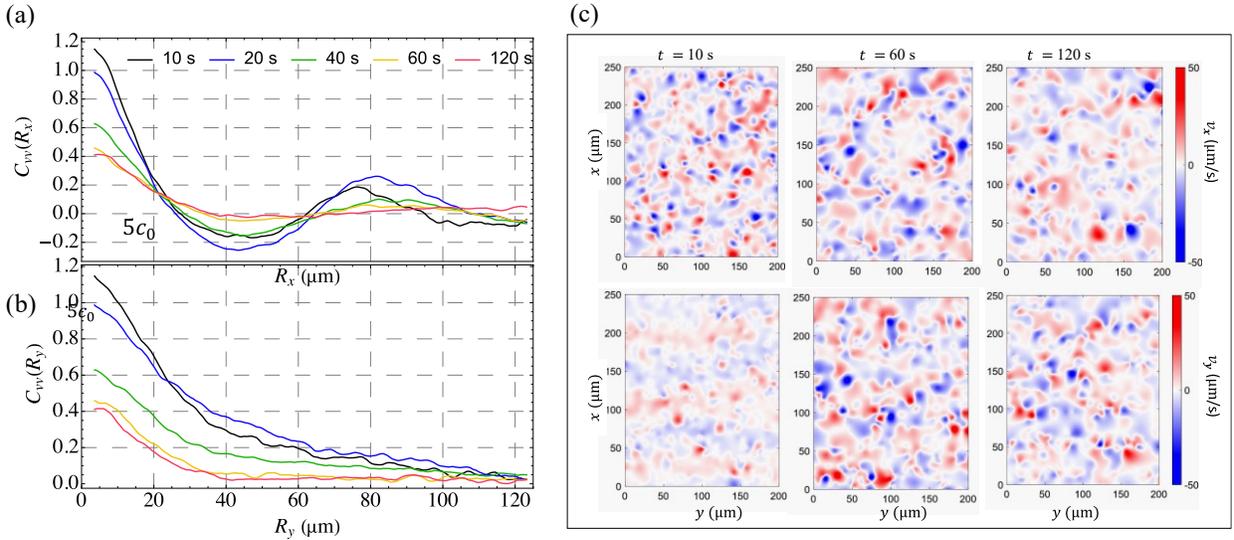

**Fig. 8.** The spatial velocity-velocity correlation function $C_{vv}(\mathbf{R})=\langle \bar{v}_{\perp y}(\mathbf{r})\bar{v}_{\perp y}(\mathbf{r}+\mathbf{R})\rangle$ of $5c_0$ calculated along the shear direction (a) $C_{vv}(R_x)$, (b) vorticity axis $C_{vv}(R_y)$ at different times $t = 10, 20, 40, 60,$ and $120$ s, and (c) snapshots of the $v_x$ and $v_y$ at different times $t = 10, 60,$ and $120$ s.

The correlations of perpendicular components of the velocities in Figs. 8(a,b) illustrate the structure of undulations well but fail to describe the turbulence. In Fig. 8(c), we show the snapshots of the $y$- and $x$-components of the total velocity field. In the undulation regime, the maps indicate active flows along the $y$-axis, which are initially weak, Fig. 8(c), and enhance as the turbulence sets in. The behavior of $x$-component is opposite, with strong flows caused by bacteria at the onset of undulations, followed by random motion at later stages. The snapshots also demonstrate that the typical scale at which the velocities reverse (which can be associated with the size of flow vortices) is about 20-35 μm, somewhat smaller than half-period $\lambda/2 \approx 40$ μm of undulations.

## 7. Energy Spectra

We quantify the energy spectra that describe spatial distributions of three quantities, the N elastic energy, kinetic energy, and enstrophy. These quantities are represented as the integrals over the area of each frame. The spectra are calculated by transforming the integrals over space coordinates into integrals over the scalar wavenumber $q$. The spectra are calculated for selected instants of time, to illustrate the evolution from uniform alignment to undulations and turbulence, Figs. 9 and 10.

**7.1.** **Elastic energy spectrum** $E_{els}(q)$ is defined through the averaged area density of the elastic energy in real space $E_{el} = \frac{K}{2L_xL_y} \int (\nabla Q_{kl}(\mathbf{r}))^2 \, d\mathbf{r} = \int E_{els}(q)dq$, where $q$ is the wavenumber, $L_xL_y$ is the frame area, and $Q_{kl}(\mathbf{r})$ represents the continuous extrapolation of the order parameter $Q_{kl}(\mathbf{r}^{(j)}) = 2n_k^{(j)}n_l^{(j)} - \delta_{kl}$ defined by the bacterial orientations $\hat{\mathbf{n}}^{(j)}$; the Frank elastic constant is estimated [39] as $K = 10$ pN. Using the discrete Fourier transform of $Q_{ij}^{(l)}$, $\tilde{Q}_{ij}(\mathbf{q}) = \frac{1}{N}\sum_{l=1}^{N} e^{-i\mathbf{q}\cdot\mathbf{r}_l} Q_{ij}^{(l)}$, we calculate the elastic energy spectrum as the weighted sum over $\mathbf{q}'$ inside an annulus of a mean radius $q$:

$$E_{els}(q) = \frac{K}{2}\sum_{\mathbf{q}'}^{w} \tilde{Q}_{ij}(\mathbf{q}')\tilde{Q}_{ij}^*(\mathbf{q}')q'^2, \tag{8}$$

where $w$ is the weighting function, derived in Supplementary Information Sections 9, 10, and Supplementary Information Fig.S7.

Figure 9(a) presents the elastic energy spectra $E_{els}(q)$ for the selected instants of time. For small $q$, corresponding to $r \approx 250 - 150$ μm, $E_{els}(q) \propto q^\nu$ at all times, as the log-log plots are linear with a slope $\nu$ increasing from 1.0 at the onset of undulations to 3.0 at well-developed turbulence. The most distinguished feature of $E_{els}(q)$ is a peak with an almost constant wavevector $q_p = 2\pi/r$ corresponding to $r \approx (70 - 80)$ μm, which grows with time. This peak is accompanied by a second-order peak, $q_{2p} = 2q_p$, that appears at later times, $t > 40$ s, Fig. 9(a). The first peak corresponds to the undulations period, $\lambda \approx$ 80 μm, Fig. 4(a). At early stages, the undulations are of a single-harmonic sinusoidal shape, and the second-order peak is absent. With time, the undulation amplitude $A$ increases, Fig. 4(b), resulting in a stronger peak at $q_p$; a somewhat higher $\lambda$ yields a slight decrease of $q_p$, Fig. 9(a). The increased $A$ reshapes the sinusoidal bend into a zigzag, causing the second-order peak $q_{2p} = 2q_p$. The development of topological turbulence at $t = 70 - 115\ s$ causes a further slight decrease of $q_p$ and broadens the small-$q$ side of the peak, so that the curve $E_{els}(q)$ has a constant slope for $q < q_p$.

The snapshots of $5c_0$ LLC at 70 s, Fig. 9(b), and 115 s, Fig. 9(c), which correspond to the saturated number of disclinations, Fig.5(e), reveal that the turbulence patterns preserve the characteristic length scale of the preceding undulations and positions of the $E_{els}(q)$ peaks. These patterns also exhibit some anisotropy, as the wave of misalignment persists along the shear (vertical) direction in Fig. 9(b,c) and Supplementary Video 6; this anisotropy correlates with the evolution of $\langle \cos 2\varphi_d \rangle$ in Fig. 5(f).

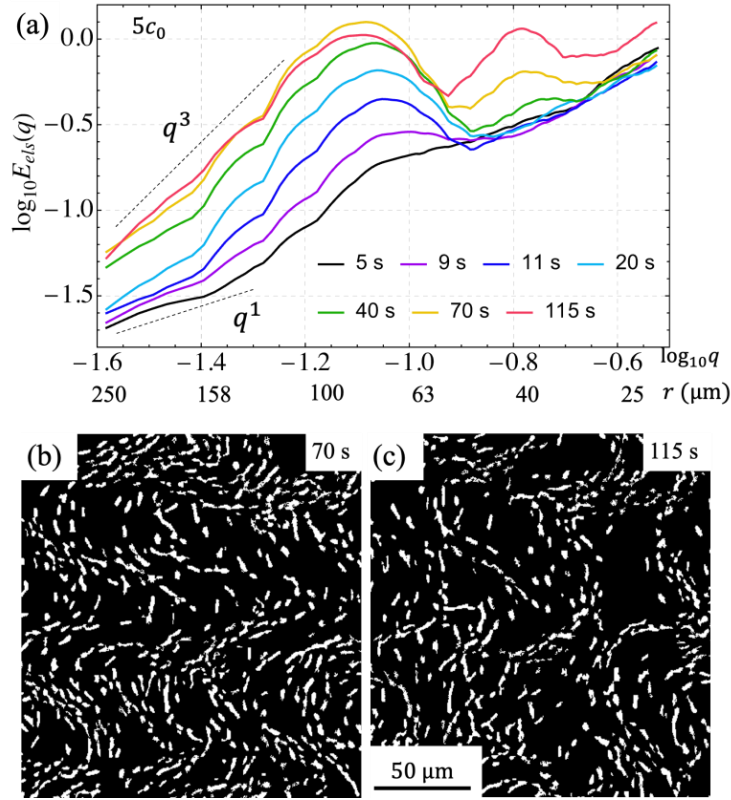

**Fig. 9.** (a) The elastic energy spectrum $E_{els}(q)$ (units N/m$^2$) as a function of the wave vector $q$ measured in μm$^{-1}$, (b) and (c) two cropped snapshots of the patterns in the $5c_0$ LLC at 70 s and 115 s, respectively, taken from Supplementary Video 6.

**7.2. Kinetic energy spectrum** $K_s(q)$ is defined through the area-averaged velocity field $\langle \mathbf{v}^2 \rangle = \frac{1}{L_x L_y} \int \mathbf{v}^2 \, d\mathbf{r} = 2 \int K_s(q) dq$. The kinetic energy spectrum is calculated similarly to the elastic energy (see Supplementary Information Section 10) as

$$K_s(q) = \sum_{\mathbf{q}'}^{w} \left( \tilde{v}_x(\mathbf{q}') \tilde{v}_x^*(\mathbf{q}') + \tilde{v}_y(\mathbf{q}') \tilde{v}_y^*(\mathbf{q}') \right). \tag{9}$$

The kinetic energy spectrum $K_s(q)$ exhibits a smooth monotonous increase with $q$ within the entire range available in the experiment, Fig.10(a). The set of quasi-linear parallel lines in the log-log plot establishes that $K_s(q) \propto q$ for all $q$ and times $t$, with two exceptions discussed later. $K_s(q) \propto q$ would be a natural result if the two-dimensional kinetic spectrum defined through $\langle \mathbf{v}^2 \rangle = 2 \int K_s^{(2)}(\mathbf{q}) d\mathbf{q}$ is homogeneous, i.e., $K_s^{(2)}(\mathbf{q}) = K_s^{(2)}$. The last equation implies $K_s(q) = 2\pi K_s^{(2)} q$. The linear dependency $K_s(q) \propto q$ implies no correlations between the velocities of neighboring bacteria. The two exceptions occur at the onset of disclination nucleation, $t \approx 20$ s, and in topological turbulence, $t \approx (60 - 120)$ s. At $t \approx 20$ s, bacteria produce alternating velocity lanes along the $y$-axis, with a period ~80 μm, Fig. 6(a). These lanes yield a strong superlinear behavior in Fig.10(a) for $t \approx 20$ s, with the slope $\nu$ increasing from 1.0 to 2.2. Furthermore, at $t \approx (60 - 120)$ s, topological turbulence produces spotty patterns of the velocity field, Fig. 6(a), and results in a weak sub-linear behavior with $\nu$ decreasing below 1.0.

The $K_s(q)$ behavior of LLC is dramatically different from a 2D dense suspension of *B. Subtilis* in an isotropic fluid, in which case $K(q) \propto q^{5/3}$ in turbulence and reaches a maximum at $q$ that corresponds to the diameter of vortices [6, 22]. This difference is caused by the elasticity of the passive N background.

**7.3. Enstrophy spectrum** $\Omega_s(q)$ is defined through the enstrophy area density $\Omega = \frac{1}{L_x L_y} \int (\text{curl } \mathbf{v})^2 \, d\mathbf{r} = \int \Omega_s(q) dq$ as

$$\Omega_s(q) = \sum_{\mathbf{q}'}^{w} \left| \tilde{v}_x(\mathbf{q}') q'_y - \tilde{v}_y(\mathbf{q}') q'_x \right|^2. \tag{10}$$

The overall behavior of $\Omega_s(q)$ in Fig.10(b) is very similar to that of the kinetic energy spectrum $K_s(q)$ in Fig.10(a) and reflects the same underlying physical effects. The only difference is in the values of the slopes, as the enstrophy spectrum generally scales as $\Omega_s(q) \propto q^{1.5}$, while at the onset of nucleation, $t \approx 20$ s, $\Omega_s(q) \propto q^3$.

The snapshots of the vorticity curl $\mathbf{v}$ in Fig. 10(c) demonstrate that the $5c_0$ LLC does not show a clear length scale. A bacterium realigns the N director parallel to its body. The realigned director field persists for some time after the bacterium leaves and serves as a "easy channel" for bacteria following the leader [42,70]. The population splits into chain-like "jets", with bacteria swimming one after the other. These jets form pairs of elongated red and blue areas in Fig. 10(c). Because of chaining, the LLC is very different from an active nematic with an isotropic environment. At $5c_0$, the nematic elasticity is still strong enough to withstand the randomizing forces of active flows.

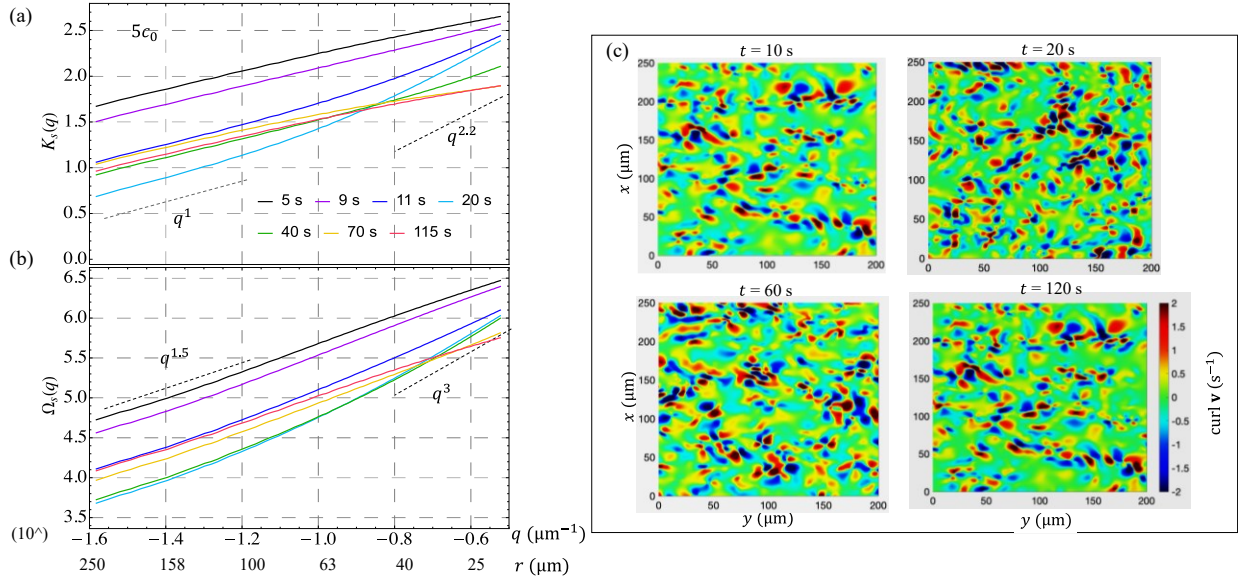

**Fig. 10.** (a) Kinetic energy spectrum $K_s(q)$ (units $(\mu m/s)^2$) and (b) enstrophy spectrum $\Omega_s(q)$ (units $s^{-2}$) as a function of the wave vector $q$ ($\mu m^{-1}$) at different times for $5c_0$. The spectra $K_s(q)$ and $\Omega_s(q)$ are averaged over 5 consecutive frames (or 1 s). (c) Vorticity curl $\mathbf{v}$ for $5c_0$ at different times.

## 8. Bacteria density modulations, clustering, and giant number fluctuation

One key feature of active matter is the presence of strong density fluctuations [6, 7, 24, 100-102]. Below, we analyze the spatial distribution and temporal evolution of the bacteria number density fluctuations.

The spatial modulation of bacteria number density is observed in the bend undulations regime. Shortly after the shear stops, $t = 3$ s, the bacteria align mostly along the shear direction; there is no clear modulation of their number density, Supplementary Information Fig. S8(a). As the amplitude $A$ of undulations increases, Fig. 4(b), the bacteria condense at the shoulders of the wave, avoiding the extrema, at which the radius of curvature is too small to accommodate an elongated bacterium, Supplementary Information Fig. S8(b). The density is modulated with a wavelength that is two times shorter than the undulation period $\lambda$. Similar density modulations have been predicted in simulations of pusher suspensions by Saintillan et al. [103] To quantify the density fluctuations evolution, we employ the Voronoi tessellation, Fig.11.

The Voronoi tessellation involves the following steps. (1) The image of each bacterium within the frame is divided into five equal parts along the long axis [104]. (2) One draws a perpendicular through the middle point of each part of the image. (3) The five middle points are connected to the similar points belonging to other bacteria in the neighborhood, and a perpendicular is drawn through the middle of each connecting segment. (4) The perpendicular lines intersect at the points that become the vertices of polygons known as the Voronoi polygons. We used the "shoelace" method to calculate the area of each polygon. For example, if $(x_i, y_i)$, $i = 1,2,3$ are the coordinates of vertices of a triangular polygon, then the area is calculated as $\sigma = (x_1 y_2 - x_2 y_1 + x_2 y_3 - x_3 y_2 + x_3 y_1 - x_1 y_3)/2$. The Voronoi polygon area corresponding to a single bacterium is found as the sum of five polygon areas corresponding to five parts of the bacterium.

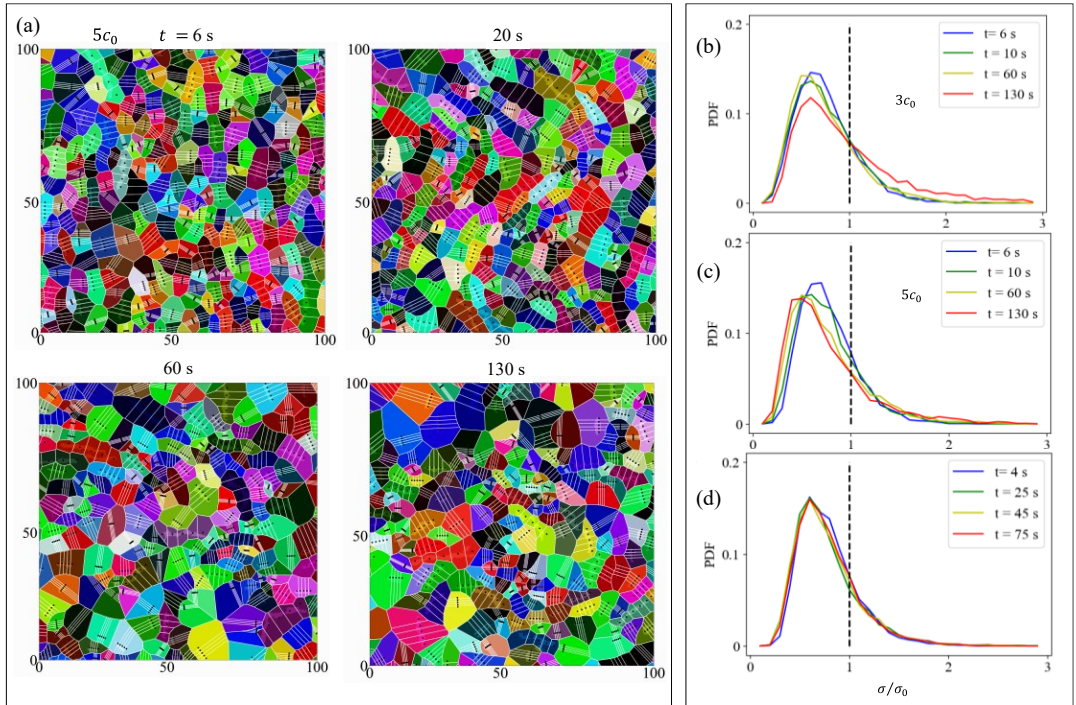

**Fig. 11.** (a) Voronoi tessellation for four instances of time, $t = 6$, 20, 60, and 130 s for a $5c_0$ LLC. (b-d) Voronoi polygons normalized area probability distribution function (PDF). (b) PDF of Voronoi polygons normalized area for $3c_0$ at $t = 6$, 20, 60, and 110 s; (c) PDF for $5c_0$ at different times of $t = 6$, 20, 60, and 130 s; and (d) PDF for $10c_0$ at different times of $t = 4$, 25, 45, and 75 s. The normalization factor $\sigma_0$ is defined as the area of the frame divided by the total number of bacteria in the frame.

Figure 11 shows the Voronoi polygons of bacteria for $5c_0$ LLC. At the early stage, $t$ = 6 s, the Voronoi polygons elongate along the shear direction; they are of approximately the same area, Fig. 11(a). Later times, $t$ = 20, 60, and 130 s, see the emergence of large areas free of bacteria (which produce large Voronoi polygons) and clusters with an elevated concentration of bacteria (small Voronoi polygons), Fig. 11(a).

The analysis of the Voronoi polygons' area distribution reveals bacterial clustering. In Fig. 11(b,c,d), the distribution of areas is shown for a normalized quantity $\sigma/\sigma_0$, where the normalization factor $\sigma_0$ is the total area of the frame divided by the number of bacteria in it. For $3c_0$, the probability distribution function (PDF) broadens with time, showing the appearance of large empty regions, Fig. 11(b). The peak of the distribution decreases and shifts to smaller areas. For $5c_0$, the shift of the distribution peak to the smaller areas is more significant, indicating stronger clustering, Fig. 11(c). For $10c_0$, the probability distribution does not change much, Fig. 11(d), since the bacterial density is already very high, so there are very few bacteria-free areas even in the well-developed turbulence.

### 8.1. Giant number fluctuations.

The density modulations in undulations and clustering in topological turbulence result in density fluctuations $\Delta N \propto \langle N \rangle^\alpha$ with $\alpha > 1/2$, which exceeds the one expected for systems at equilibrium, $\Delta N \propto \langle N \rangle^{1/2}$, where $\langle N \rangle$ is the average number of particles and $\Delta N = \sqrt{\langle (N - \langle N \rangle)^2 \rangle}$ is the standard deviation. For $3c_0$, α increases from 0.72 to 0.88 over time, Fig.12. These values are similar to ones found in other systems. Nishiguchi et al. measured $\alpha \approx 0.63$ for a 2D system of *B. Subtilis* in an isotropic fluid [24], while Liu et al. measured $\alpha$ to increase from $\alpha \approx 0.6$ to $\alpha \approx 0.83$ over 120 s in an active turbulent 3D system of *E. coli*, also in an isotropic fluid [25]. Different models predict different limits for α [31, 105-107]. Recent simulations of an active nematic with multiparticle collision dynamics, which resembles an LLC, show that α can grow from $1/2$ at low activities to 1 at intermediate activities and then decreases to about 0.8 at the highest activities, which diminish the degree of the nematic order [108].

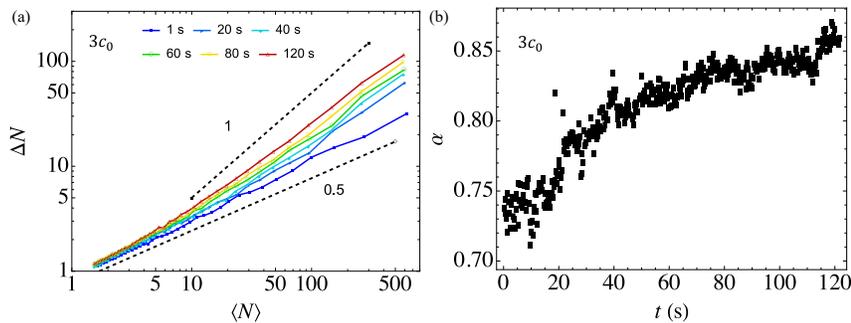

**Fig.12.** (a) Standard deviation $\Delta N$ of the number of bacteria as a function of the average number $\langle N \rangle$ of bacteria at different times for $3c_0$; (b) the fitting exponent α of $\Delta N \propto \langle N \rangle^\alpha$ as a function of time.

## Conclusions

Flagellated microswimmers *B. Subtilis* dispersed in a nematic lyotropic chromonic liquid crystal DSCG form a living liquid crystal (LLC). We explore the emergence and development of topological turbulence from a uniformly aligned state produced by shear of a thin (10 μm) layer of LLC. The experimental setup has a large area to avoid the effect of lateral confinement and eliminates in-plane surface anchoring, which could impose a permanent anisotropy on the system. Upon shear cessation, bacterial activity destabilizes the uniform alignment, producing self-amplifying bend undulations, with an amplitude that initially grows linearly as a function of time, then saturates, while the period λ remains almost constant. The active stress that drives the growth of amplitude is released via nucleating $\pm 1/2$ defect pairs at the extrema of the bend. The defects multiply until the undulation pattern is lost and a regime of topological turbulence is established. Increasing the bacteria concentration results in a smaller λ, a higher growth rate of undulations, and an increased number of the $\pm 1/2$ defects.

To identify the disclinations and analyze their dynamics in the bacterial population, we propose an approach based on the Delaunay triangulation mesh. The disclinations of strength $\pm 1/2$ appear as bound pairs (BP) at the onset of nucleation and develop into isolated disclinations at later stages. The isolated +1/2 disclinations and $\pm 1/2$ pairs are initially well aligned along the $y$-axis orthogonal to the shear direction, Fig.5(f). As time progresses, the number densities of disclinations grow and saturate. The orientational diffusion of **p** is much faster than the translational diffusion of +1/2 cores along the vertical $x$-axis, as shown in Fig. 5(h).

The number density of disclinations in the strongly active $10c_0$ LLC shows a nonmonotonic time dependence, Fig. 5(e), caused by the "alignment and activity-triggered annihilation." Namely, a fast-moving +1/2 disclination of one nucleated pair reaches a -1/2 disclination of a similarly aligned neighbouring pair and annihilates with it, thus reducing the total number of defects.

The velocities of individual bacteria reveal the presence of two components: the actual individual swimming velocity and the velocity of an underlying active flow caused by bacterial interactions. Immediately after the shear cessation, only the first component is present, with a maximum speed ~20 μm/s of bacteria swimming along the shear direction. Emerging bend undulations produce active flows orthogonal to the shear direction. As a result, the bacterial velocity makes a significant angle with the body axis. At the transition from bend to turbulence, most bacteria realign along the active flow. The active flows result from hydrodynamic interactions of the bacteria. Each bacterium produces a pair of axial forces, $\mathbf{F}^{(i)}_{prop} = -\mathbf{F}^{(i)}_{drag}$. The active flow could result from adding $\mathbf{F}^{(i)}_{prop}$ of a bacterium "$i$" and $\mathbf{F}^{(j)}_{prop}$ of a bacterium "$j$" or adding $\mathbf{F}^{(i)}_{drag}$ and $\mathbf{F}^{(j)}_{prop}$, or $\mathbf{F}^{(i)}_{drag}$ and $\mathbf{F}^{(j)}_{drag}$, Fig.6(b).

The calculated density-density, orientation-orientation, and velocity-velocity correlation functions provide additional information on LLC behavior. The positive maximum of density-density correlations at the typical length of the bacterium body illustrates the tendency of bacteria to form chains; a leading bacterium leaves an easy passage in the N director field for other bacteria to follow. The orientation-orientation correlations show maxima at $r \approx 40$ μm and $\approx 80$ μm, which correspond to the wave of undulations; these typical correlation distances are traceable also in the turbulence. The velocity-velocity correlation along the shear direction shows a negative minimum at $R_x \approx 40$ μm and a maximum at $R_x \approx 80$ μm in the bend instabilities regime, Fig.8. The negative minimum is caused by antiparallel active flows that develop at the two neighboring bend extrema, while the maximum corresponds to the full wavelength $\lambda \approx 80$ μm of undulations, Fig.6.

The elastic energy spectra exhibit a peak with an almost constant wavevector that corresponds to the period of undulations. When the sinusoidal undulations transform into zigzag patterns, a second-order peak appears at a doubled value of the wavevector, Fig. 9(a). The kinetic energy and enstrophy spectra associated with the velocity fields exhibit a monotonous increase with the wavevector within the entire range available in the experiment. The kinetic energy spectrum grows linearly with the wavevector, except at the onset of nucleation and in the topological turbulence. In the latter case, one observes a sublinear behavior that is dramatically different from the active nematic formed by *B. Subtilis* swimming in an isotropic environment. Also different are the patterns of vorticity, which in the LLCs do not show a clear length scale, in contrast to their isotropic counterparts. The LLC vorticity is affected by the chaining of swimming bacteria.

Besides the differences in the energy and enstrophy spectra, orientational order of disclinations, chaining of bacteria, there are other unique dynamic properties of the LLC that distinguish them from active nematics in which the self-propelled units are in an isotropic Newtonian environment. A long memory of the imposed macroscopic orientational order is one example; an intriguing, related effect is that in diluted dispersions, $c = (0.5 - 2)c_0$, the bacteria continue to swim up and down the pre-imposed shear direction for a very long time, about 3 min after the cessation of flow, Supplementary Fig. S3. In diluted LLCs, the separation of bacteria is larger than the range of their hydrodynamic interactions, which explains the long-lasting orientational order.

To conclude, we analyzed the LLCs in which the active nematic part, the dispersion of swimming bacteria, is controlled by an orientationally ordered background of the passive nematic, representing a lyotropic chromonic liquid crystal DSCG. The 10 μm scale of the bacterial bodies allows us to image and analyze the relevant parameters with great detail by optical microscopy. The tangential anchoring of the passive nematic at the bacterial bodies allows us to connect the passive director field to the bacterial orientations and to calculate important parameters characterizing the cascade of orientational disorder, first through bend undulations and then through nucleation of disclinations and turbulence. Combined with the previously determined material properties of DSCG, such as elastic constants and viscosities [81,82], resolved for splay, bend, and twist of the director, DSCG behavior in confinement [80,109-111], the results complete a comprehensive description of the experimentally assessable type of active matter in which the self-propelled microswimmers move in a viscoelastic orientationally ordered medium.

## Conflicts of interest

There are no conflicts to declare.

## Data Availability

Data for this article, including binary videos for the experimental data and the codes used for the analysis are available at this [GitHub repository](#).

## Acknowledgments.

The authors thank L. Reichel and A. Buccini for their valuable discussions and insightful comments. The work was supported by NSF grants DMS-1729509 and DMR-2341830.

## References.

# Supplemental Information

# Bend Instabilities and Topological Turbulence in Shear-Aligned Living Liquid Crystal

Hend Baza[1,2], Fei Chen[3], Taras Turiv[1,4], Sergij V. Shiyanovskii[1,4], Oleg D. Lavrentovich[1,2,4*]

[1]Advanced Materials and Liquid Crystal Institute, Kent State University, Kent, OH 44242, USA
[2]Department of Physics, Kent State University, Kent, OH 44242, USA
[3]Department of Mathematical Sciences, Kent State University, Kent, OH 44242, USA
[4]Materials Science Graduate Program, Kent State University, Kent, OH 44242, USA
* Corresponding author. Email: olavrent@kent.edu

Keywords: active nematics, active matter, living liquid crystal, shearing, rheology, topological defects, topological turbulence, Undulations.

## 1. Alignment of bacteria during the shear.

At shear rates $\dot{\gamma} \leq 1 \text{s}^{-1}$, LCLC was reported to align along the vorticity direction [1]. Similarly, the nematic director of LN and bacteria align along the vorticity axis at $\dot{\gamma} \leq 1 \text{s}^{-1}$, Fig. S1 (a). Higher shear rates $1 \leq \dot{\gamma} \leq 500 \text{ s}^{-1}$ result in a polydomain structure in both passive N and LN, Fig. S1 (b). At higher shear rates $\dot{\gamma} = 1000 \text{ s}^{-1}$, bacteria align along the shear direction, while the director shows small periodic tilts to the left and right of the shear direction.

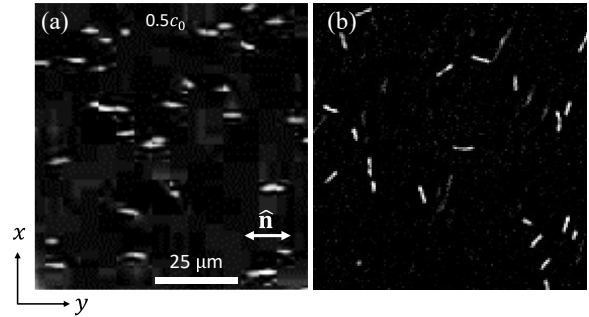

***Fig. S1.*** *(a) Shear at $\dot{\gamma} = 1 \, s^{-1}$ aligns the director and the bacteria along the vorticity y-axis (log-rolling regime); (b) shear at $\dot{\gamma} = 15 \, s^{-1}$ produces polydomain texture. 13 wt.% of DSCG in TB, 25°C; shear cell gap 10 μm. Bacterial concentration $0.5c_0$.*

## 2. Relaxation of an inactive N after shear cessation.

Once the shear at $\dot{\gamma} = 1000$ s$^{-1}$ stops at $t = 0$ s, the passive N develops stripes parallel to the shear $x$-direction, which at $t \approx 1$ s are replaced by bands perpendicular to it, Fig. S2(a) and Supplementary Video 1. The wavelength $\lambda_0$ of the N director bend increases with time, as discussed in the main text. The bands quickly develop into a polydomain texture, at $t \approx 3$ s . This behavior is different from the relaxation of the LN, illustrated in Fig.S2(b), in which case the period of band wave increases with time only slightly.

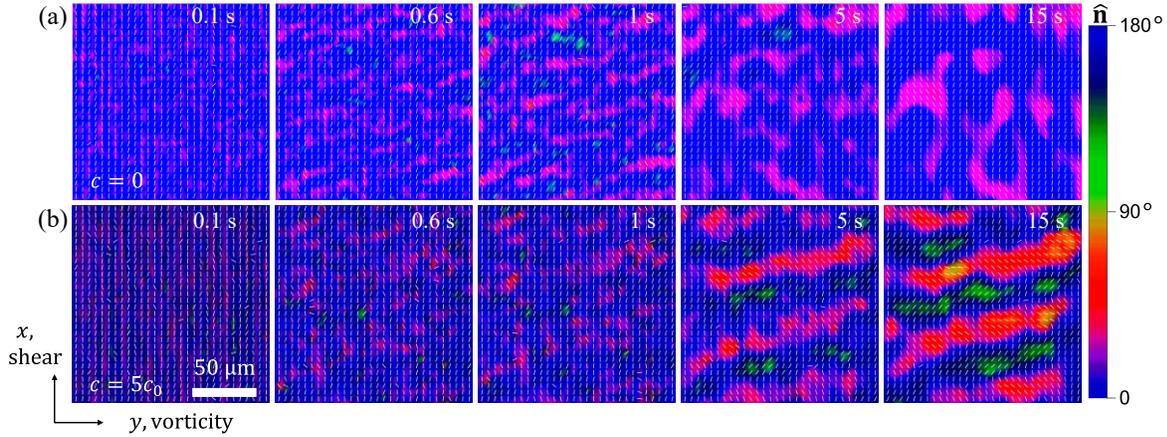

***Fig. S2.*** *PPM textures of (a) a passive N, $c = 0$, and (b) an LN, $c = 5c_0$; the material is shear-aligned at $t = -0.1$ s and relaxes after the cessation of flow at $t = 0$. The N director $\hat{\mathbf{n}}$ is shown by white bars and pseudocolors. The cell thickness is $h = 10$ μm. See Supplementary Video 1 for (a) and Supplementary Video 2 for (b).*

## 3. Alignment of bacteria after shear cessation.

All experiments of the LN relaxation are performed after the samples were sheared at $\dot{\gamma} = 1000$ s$^{-1}$ for 30 s, so that all the bacteria are aligned along the $x$-axis of shear direction. At small concentrations of bacteria, (0.5-2) $c_0$, the cessation of shear does not cause a significant change of bacterial alignment: the bacteria continue to swim along the shear direction with equal probability to swim up and down; this aligned pattern persists for ~ 3 minutes, Fig. S3.

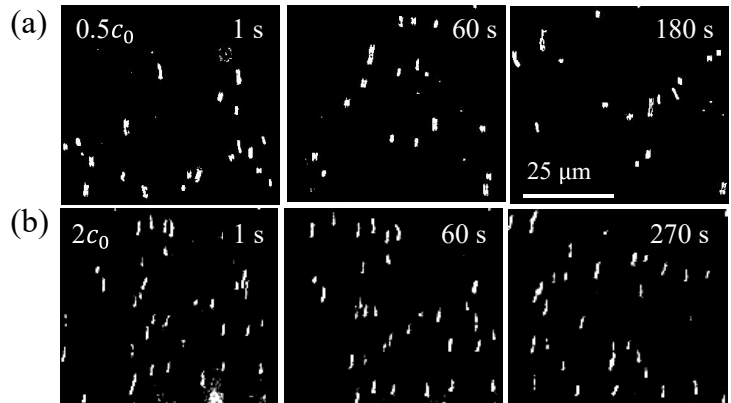

***Fig. S3.*** *Bacteria remain aligned along the shear direction in (a) $0.5c_0$ LN and (b) $2c_0$ LN. DSCG 13 wt.% in TB; 25 °C; shear cell gap 10 μm. The initial shear with the rate $\dot{\gamma} = 1000$ s$^{-1}$ for 30 s.*

### 4. Video Processing.

The frames within each video are processed to enhance the contrast of the bacteria. Our method includes three main steps:

***Step 1: Colored to Grayscale images preparation.*** The digital-colored image collected using POM without polarizers, equipped with Am-Scope camera, 1216×1824 pixels with a frame rate of 5 f/s and magnification of 20× such that each pixel is 0.4 μm in length, consists of three-color channels, namely red, green, and blue (RGB). We split the channels using ImageJ and selected the green channel, which provides the highest contrast between the bacteria and the background. Each pixel value is between 0 to 255, with 0 being the darkest and 255 the brightest.

***Step 2: Background Removal.*** To identify the individual bacteria, we remove the background and other undesired particles such as dust, whose pixel values are usually below 80. Hence, we first set all pixels values that are below 80 to 0. The pixels containing bacteria are brighter than the background pixels. For each pixel, we calculate its temporal median μ and temporal standard deviation σ. Since the illumination also varies along time dimension, at each pixel, we treat any value that is less than $\mu - \sigma$ as background and those above as bacteria. We convert the grayscale images to binary images by assigning the background a brightness of 0, while pixels with bacteria are assigned a brightness 1.

***Step 3: Denoising.*** The area of most bacteria is much larger than 6 pixels. Hence, objects formed by 6 or fewer connected pixels are considered as "noise" and removed.

### 5. Local director of bacteria.

We use MatLab built-in "regionprops" function to extract the center of mass, the long and short axes, and the orientation of bacteria. A rod-like $j^{th}$ bacterium makes an angle $\theta_j \in (-\frac{\pi}{2}, \frac{\pi}{2}]$ with the shear direction.

### 6. Undulations wavelength and amplitude analysis.

The distribution of bacterial orientation evolves over time. Immediately after the shear cessation, almost all bacteria align along the shear direction. As time progresses, a bend undulation instability emerges, Fig. S4.

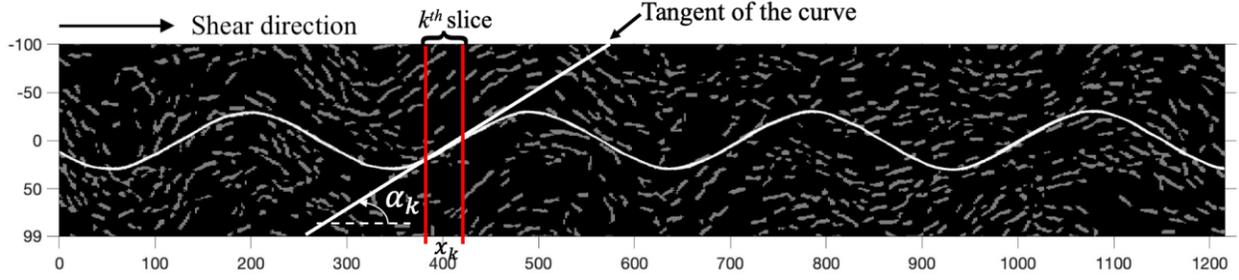

**Fig. S4.** *Analysis of bend instability in $5c_0$ dispersion. The undulation wave is fitted for slices of a height 200 pixels (80 μm) and width 32 pixels (12.8 μm); the red lines show the edges of one of these slices.*

To characterize the undulation wave, we divide the frame into slices of height 200 pixels (80 μm) and width 32 pixels (12.8 μm), Fig. S4. Within $k^{th}$ slice, we average the orientation of bacteria to obtain the tilt angle $\bar{\alpha}_k$. Let $x_k$ be the $x$-coordinate of the center of the $k^{th}$ slice. A set of experimental values $\alpha_k := \alpha(x_k)$ are fitted such that $\sum_k \| e^{2i\bar{\alpha}_k} - e^{2i\alpha_k} \|^2$ is minimized, where $\alpha_k$ is the tangent angle of the curve to be fitted at position $x_k$. Analytically

$$\alpha(x) = \alpha_{max} sn(J(x-h), m), \tag{S1}$$

describes the development of undulations in the passive smectic liquid crystal [2], where $sn(\cdot)$ is the Jacobi elliptic function [3], $J > 0$ is the Jacobi amplitude, $0 < m < 1$ is the elliptic modulus, $h$ is the phase shift along the shear direction, and $\alpha_{max}$ is the maximum tilt angle of the wave [2]. The fitting results in the values of $\alpha_{max}$, $J$, $h$, and $m$. For small $m$, Eq. (S1) describes a shape close to a sinusoid, so that $J \approx \frac{2\pi}{\lambda}$, where $\lambda$ is the period of undulations. As $m$, which is an analog of time in our case, increases, the sinusoid gradually transforms into a saw-tooth profile, with sharp realignment at the extrema of the wave. The displacement $u$ of the wave at position $x$ obtained by integrating Eq. (S1),

$$u(x) = \frac{\alpha_{max}}{J\sqrt{m}} \log\left[dn(J(l-h)|m) - \sqrt{m}\, cn(J(l-h)|m)\right] + C, \tag{S2}$$

allows one to calculate the amplitude as $A = \frac{(u_{max}-u_{min})}{2}$ and the period $\lambda$ as the distance between the two maxima of $u(x)$; here $C$ is an arbitrary constant, $cn(\cdot)$ and $dn(\cdot)$ are the Jacobi elliptic function [3].

## 7. Evolution of disclinations.

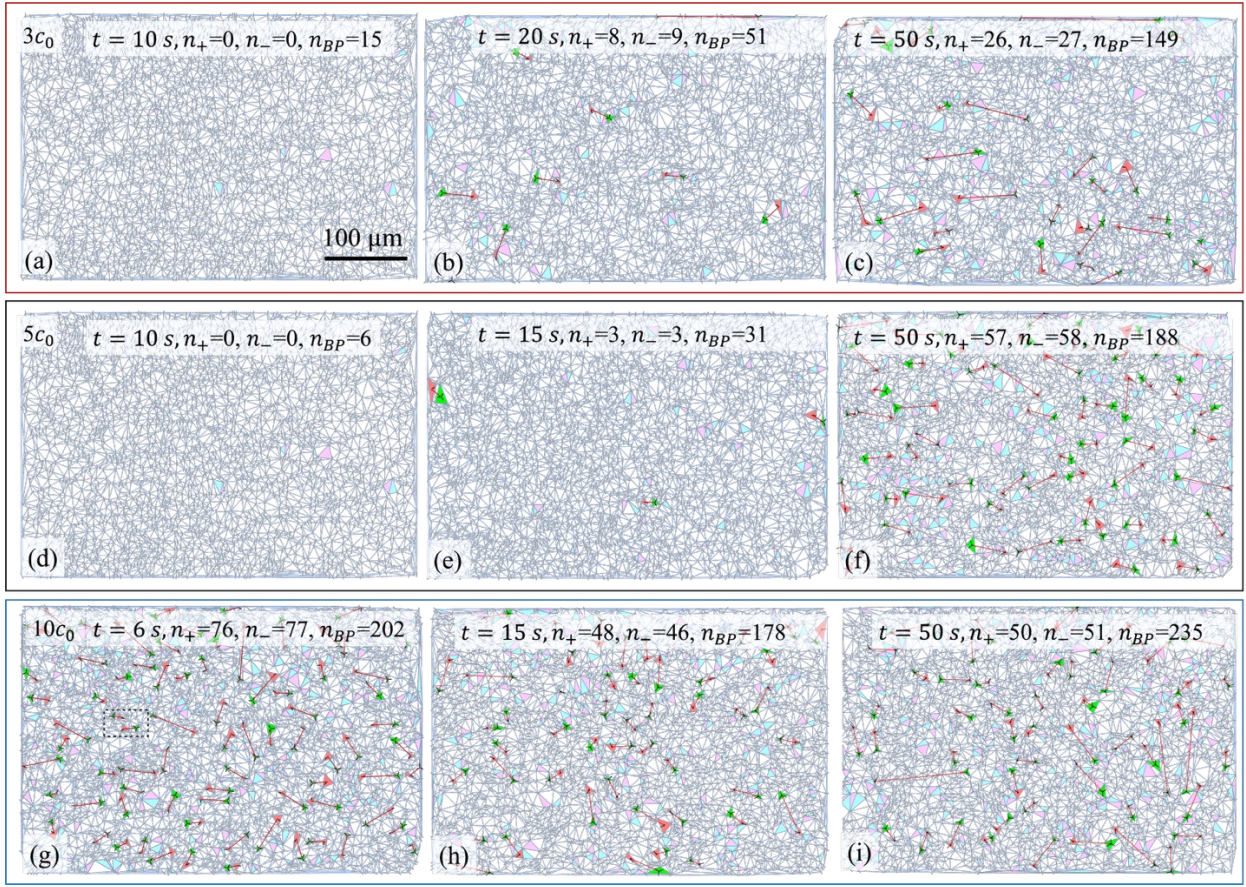

**Fig. S5.** *Evolution of disclinations in LN with $c = 3c_0$ (a,b,c), $5c_0$ (d,e,f), and $10c_0$ (g,h,i). Each frame contains $n_+$ red triangles (with $+1/2$ disclinations inside), $n_-$ green ones (with $-1/2$ disclinations), and $n_{BP}$ BPs composed of magenta ($+1/2$) and cyan ($-1/2$) triangles; enlarged part of frame (g) is presented as Fig.6(d) in the Main text.*

## 8. Bacteria tracking algorithm.

A bacterium is identified as a set of connected bright pixels with an area $\geq 16$ pixels ($2.56$ μm$^2$) or with an aspect ratio (length/width) $\geq 1.5$. To track a bacterium, we overlap two consecutive frames $n$ and $n+1$ (separated by 0.2 s) in one image, which shows that a bacterium in the second frame overlaps with some part of its body in the first. The tracking scheme is as follows:

i. Find $\mathbf{r}_n^{(j)}$, the location of the center of mass of the $j^{th}$ bacterium in the current frame $n$.

ii. Find $\mathbf{r}_{n+1}^{(l)}$, the location of the center of mass a bacterium in $n+1$ that is the closest to $\mathbf{r}_n^{(j)}$.

iii. Compare the pixel lists that contain $\mathbf{r}_n^{(j)}$, in $n$ and $\mathbf{r}_{n+1}^{(l)}$ in $n+1$. If the two lists share at least one pixel, $\mathbf{r}_n^{(j)}$ and $\mathbf{r}_{n+1}^{(l)}$ represent the center of mass of the same bacterium $\mathbf{r}_{n+1}^{(l)} = \mathbf{r}_{n+1}^{(j)}$. Otherwise, we move to the second-closest mass center in $n+1$ to $\mathbf{r}_n^{(j)}$ and repeat this step.

iv. If we cannot find an $\mathbf{r}_{n+1}^{(l)}$ that represent the same bacteria for $\mathbf{r}_n^{(j)}$, we claim that the corresponding bacterium in $n$ moves out of the frame $n+1$.

## 9. *Velocities of bacteria.*

Using bacteria tracking, we can calculate their velocities since the time $\Delta t$ elapsed between two consecutive frames is known to be 0.2 s. To increase the accuracy, for each bacterium in any three consecutive frames $n-1$, $n$, and $n+1$, we calculate its velocity at the time of the middle frame by dividing the distance its mass center has traveled by 0.4 seconds.

$$\mathbf{v}_n^{(j)} = \left[\left(\mathbf{r}_n^{(j)} - \mathbf{r}_{n-1}^{(j)}\right) + \left(\mathbf{r}_{n+1}^{(j)} - \mathbf{r}_n^{(j)}\right)\right]/2\Delta t. \tag{S3}$$

Let $\mathbf{v}_n^{(j)}$ and $\theta_n^{(j)}$ be the velocity and orientation of the $j^{th}$ bacterium in the frame $n$, respectively. In what follows, we drop the subscript "$n$" to simplify the notations. In the absence of other bacteria, $\mathbf{v}^{(j)}$ is collinear with the director $\hat{\mathbf{n}}^{(j)} = \left(\cos\theta_n^{(j)}, \sin\theta_n^{(j)}\right)^T$, where $T$ denotes the transpose of a row vector. However, the experimental $\mathbf{v}^{(j)}$ often deviates from $\hat{\mathbf{n}}^{(j)}$; see, for example, Fig. S6. We decompose $\mathbf{v}^{(j)}$ into a component $\mathbf{v}_\parallel^{(j)}$ parallel to the bacterial body and $\mathbf{v}_\perp^{(j)}$ perpendicular to it, such that $\mathbf{v}_\parallel^{(j)} = \left(\mathbf{v}^{(j)} \cdot \hat{\mathbf{n}}^{(j)}\right)\hat{\mathbf{n}}^{(j)}$, and

$$\mathbf{v}_\perp^{(j)} = \mathbf{v}^{(j)} - \mathbf{v}_\parallel^{(j)}. \tag{S4}$$

A non-vanishing $\mathbf{v}_\perp^{(j)}$ is attributed to the active flows resulting from bacterial interactions; these flows advect a bacterium along a direction different from $\hat{\mathbf{n}}^{(j)}$. The observed velocity of a $j^{th}$ bacterium advected by an active flow can be represented as a sum

$$\mathbf{v}^{(j)} = \mathbf{v}_s^{(j)} + \mathbf{v}_f^{(j)}, \tag{S5}$$

where $\mathbf{v}_s^{(j)}$ is the swimming velocity and $\mathbf{v}_f^{(j)}$ is the active flow velocity at the location of the bacterium. Since a bacterium swims along its long axis, $\mathbf{v}_s^{(j)} = \left(\mathbf{v}_s^{(j)} \cdot \hat{\mathbf{n}}^{(j)}\right) \hat{\mathbf{n}}^{(j)}$, while $\mathbf{v}_\perp^{(j)}$ is attributed to $\mathbf{v}_f^{(j)}$:

$$\mathbf{v}_\perp^{(j)} = \mathbf{v}_f^{(j)} - \left(\mathbf{v}_f^{(j)} \cdot \hat{\mathbf{n}}^{(j)}\right) \hat{\mathbf{n}}^{(j)}, \tag{S6}$$

Since $\mathbf{v}_\perp^{(j)}$ can be calculated from Eq. (S6), the only unknown in Eq. (S6) is the local flow velocity $\mathbf{v}_f^{(j)}$, which we represent as $\mathbf{v}_f^{(j)} = (v_x, v_y)^T$ and rewrite Eq. (S6) as

$$\begin{bmatrix} 1 - \cos^2\theta_n^{(j)} & \cos\theta_n^{(j)} \sin\theta_n^{(j)} \\ \cos\theta_n^{(j)} \sin\theta_n^{(j)} & 1 - \sin^2\theta_n^{(j)} \end{bmatrix} \begin{pmatrix} v_x \\ v_y \end{pmatrix} = \mathbf{v}_\perp^{(j)}. \tag{S7}$$

The matrix in Eq. (S7) is singular since its determinant equals 0. To accurately estimate $\mathbf{v}_f^{(j)}$, we consider a number of k ≥ 2 bacteria inside the circle of the radius r centered at the mass center of the $j^{th}$ bacterium. The radius r is chosen by the following two principles:

i. The flow velocities inside the circle can be considered the same. We choose r = 48 pixels (19.2 μm)
ii. The circle contains enough bacteria with different orientations. Since the bacteria in the frame $n$ are sparse, we also consider the same circle in the frames $n - 1$ and $n + 1$, in order to increase the number k of bacteria. Then the corresponding coefficient matrix of Eq. (S7) is not ill-conditioned but overdetermined. Therefore, the least-squares solution is a proper approximation of the local flow velocity.

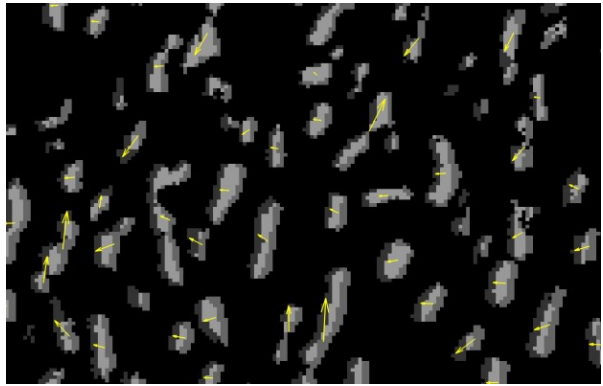

*Fig. S6. The overlapped images of two consecutive frames $n$ and $n + 1$. The brightest grey represents the overlapped part of the bacteria in both frames. The medium grey represents the part of the bacteria in the frame $n$. The darkest grey represents the part of bacteria in the frame $n + 1$. Yellow arrows pointing from the mass centers in the frame $n$ show the velocity of bacteria; their*

*length is proportional to the speed. Bright areas smaller than* 16 *pixels or with an aspect ratio less than* 1.5 *are ignored.*

An example of bacterial velocity field is presented in Fig.S7. The flows along the $y$-axis in the bend undulations are followed by chaotic flows in topological turbulence, Fig.7.

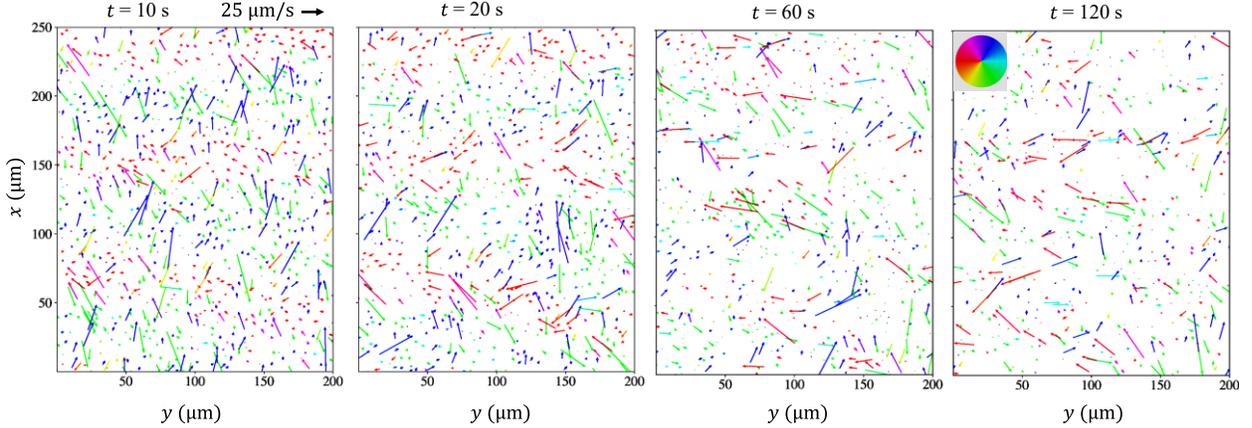

***Fig. S7.*** *The individual velocities* $\mathbf{v}^{(j)}$ *of the bacteria in* $5c_0$ *LN are shown as colored arrows. The colors indicate the velocity direction according to the color wheel. The frames at $t =$10 s (onset of weak bend instability), 20 s (well-developed bend instability), 60 s (transition from bend instability to topological turbulence), and 120 s (topological turbulence). The arrow on the top shows the velocity scale, 25 µm/s.*

## 10. Correlation functions.

To evaluate correlation functions from the images we introduce the sets of circles of radius $r$ and width $2\delta$ centered at $r_l$, the mass center of the $l^{th}$ bacterium and calculate the weighted number of bacteria in these rings. The weight function for a $k^{th}$ bacterium at a distance $r_{lk}$ from the $l^{th}$ bacterium in this ring

$$w(r, r_{lk}) = \begin{cases} 1 - (r_{lk} - r)^2/\delta^2, & \text{if } |r_{lk} - r| < \delta; \\ 0, & \text{if } |r_{lk} - r| \geq \delta \end{cases} \quad (S8)$$

allows us to reconstruct continuous $r$-dependence of correlation functions with statistically averaged values.

### 10.1. <u>Density-density correlation function.</u>

The density-density isotropic correlation function is

$$C_{00}^{(0)}(r) = \frac{1}{W(r)} \sum_{l=1}^{N_r} \sum_{k=1}^{N_{r,l}} w(r, r_{lk}) - 1, \tag{S9}$$

where $N_r$ is the total number of rings of a radius $r$, $W(r) = \sum_{l=1}^{N_r} W_l(r)$ is the total expected weighted number of bacteria in the rings of a radius $r$, $W_l(r)$ is the expected weighted number of bacteria in the ring of a radius $r$ around the $l^{th}$ bacterium. If the ring is complete, $W_l(r) = 2\pi\rho \int_{r-\delta}^{r+\delta} w(r, r_{lk}) r_{lk} dr_{lk} = \frac{8}{3}\pi\rho r\delta$, otherwise the integral is taken over the areas marked yellow in Fig. S8. Here $\rho$ is the average number density of bacteria in a frame.

In a similar way, we also construct the anisotropic density-density correlation function

$$C_{00}^{(2)}(r) = \frac{1}{W(r)} \sum_{l=1}^{N_r} \sum_{k=1}^{N} \cos(2\varphi_k) w(r, r_{lk}), \tag{S10}$$

where $\varphi_k \in (-\pi, \pi]$ is the angle between the line connecting $l^{th}$ and $k^{th}$ bacteria and the shear direction, Fig.1(c).

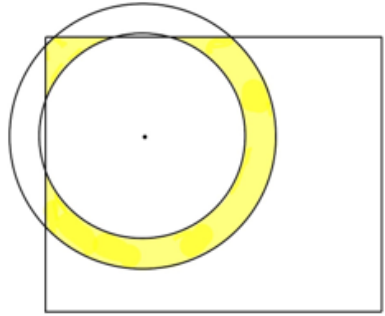

***Fig. S8.*** *A partial ring with a radius $r$ near the edge of a frame.*

### 10.2. <u>Orientation-Orientation Correlation Function.</u>

The average isotropic orientation-orientation correlation function $C_{22}^{(0)}(r)$

$$C_{22}^{(0)}(r) = \frac{1}{W(r)} \sum_{l=1}^{N_r} \sum_{k=1}^{N} \cos 2(\theta_l - \theta_k) w(r, r_{lk}) - S_N^2, \tag{S11}$$

where $S_N$ is the scalar order parameter measured for all $N$ bacteria in the rings of a radius $r$ in the eight consecutive frames, and $\theta_l, \theta_k \in (-\pi, \pi]$ are the orientations of the $l^{th}$ and $k^{th}$ bacteria, respectively.

The anisotropic orientation-orientation correlation function $C_{22}^{(2)}(r)$ is calculated as

$$C_{22}^{(2)}(r) = \frac{1}{W(r)} \sum_{l=1}^{N_r} \sum_{k=1}^{N} \cos(2\varphi_k) \cos 2(\theta_l - \theta_k) w(r, r_{lk}), \tag{S12}$$

### 10.3. <u>Velocity Correlation Function.</u>

The velocity-velocity correlation function is defined as

$$\langle \mathbf{v}_l \cdot \mathbf{v}_k \rangle(r) = \frac{1}{W(r)} \sum_{l=1}^{N_r} \sum_{k=1}^{N} \mathbf{v}_l \cdot \mathbf{v}_k \, w(r, r_{lk}), \tag{S13}$$

where $\mathbf{v}_l$ is the velocity of the $l^{th}$ bacterium, $\mathbf{v}_k$ is the velocity of the $k^{th}$ bacteria in the ring centered at $r_l$, and the brackets denote the average over $l$ and $k$.

## 11. Spectral analysis
### 11.1. *Elastic Energy Spectrum.*

The orientation of the $l$-th bacterium is defined by the unit vector $\hat{\mathbf{n}}^{(l)} = \left(n_x^{(l)}, n_y^{(l)}\right)$ along its long axis. We derive the elastic energy of the system using the Q-tensor of $N$ individual bacteria $Q_{ij}^{(l)} = 2n_i^{(l)} n_j^{(l)} - \delta_{ij}$, $l = 1, \ldots, N$ and its discrete Fourier transform (DFT) is given by

$$\tilde{Q}_{ij}(\mathbf{q}) = \frac{1}{N} \sum_{l=1}^{N} e^{-i\mathbf{q}\cdot\mathbf{r}_l} Q_{ij}^{(l)}, \tag{S14}$$

where $\mathbf{q} = \left(\frac{2\pi}{L_x} k_x, \frac{2\pi}{L_y} k_y\right)$, $\tilde{Q}_{ij}(-\mathbf{q}) = \tilde{Q}_{ij}^*(\mathbf{q})$, and $\mathbf{r}_l$ is the location of the $l$-th bacterium.

$k_x = \left(-k_x^{(m)}, \ldots, k_x^{(m)}\right), k_y = \left(-k_y^{(m)}, \ldots, k_y^{(m)}\right), \overline{N} = \left(2k_x^{(m)} + 1\right)\left(2k_y^{(m)} + 1\right) \leq N.$

where $k_x^{(m)}$ is the maximum value of the harmonic order $k_x$, $k_y^{(m)}$ is a similar quantity for $k_y$; $\overline{N}$ is the total number of Fourier harmonics, which should not be larger than the number of bacteria $N$. The inverse DFT is

$$\overline{Q}_{ij}^{(l)} = \sum_{k_x=-k_x^{(m)}}^{k_x^{(m)}} \sum_{k_y=-k_y^{(m)}}^{k_y^{(m)}} \tilde{Q}_{ij}(\mathbf{q}) e^{i\mathbf{q}\cdot\mathbf{r}_l} = \sum_q \tilde{Q}_{ij}(\mathbf{q}) e^{i\mathbf{q}\cdot\mathbf{r}_l}, \tag{S15}$$

Note that $\overline{Q}_{ij}^{(l)} = Q_{ij}^{(l)}$ if the number of Fourier harmonics $\overline{N}$ is equal to the number of bacteria $N$. Continuous extension $\mathbf{r}_l \to \mathbf{r}$ and $\overline{Q}_{ij}^{(l)} \to Q_{ij}(\mathbf{r})$, where

$$Q_{ij}(\mathbf{r}) = \sum_q \tilde{Q}_{ij}(\mathbf{q}) e^{i\mathbf{q}\cdot\mathbf{r}} = \sum_{k_x=-k_x^{(m)}}^{k_x^{(m)}} \sum_{k_y=-k_y^{(m)}}^{k_y^{(m)}} e^{i\mathbf{q}\cdot\mathbf{r}} \tilde{Q}_{ij}(\mathbf{q}) \tag{S16}$$

results in the averaged area density of the elastic energy

$$E_{el} = \frac{K}{2L_x L_y} \int \left(\nabla Q_{ij}(\mathbf{r})\right)^2 d\mathbf{r}. \tag{S17}$$

Using Eq. (S16) and orthogonality of the Fourier harmonics, one finds

$$E_{el} = \frac{K}{2} \sum_q \tilde{Q}_{ij}(\mathbf{q}) \tilde{Q}_{ij}^*(\mathbf{q}) q^2 = \frac{K}{2} \sum_{k_x=-k_x^{(m)}}^{k_x^{(m)}} \sum_{k_y=-k_y^{(m)}}^{k_y^{(m)}} \tilde{Q}_{ij}(\mathbf{q}) \tilde{Q}_{ij}^*(\mathbf{q}) q^2 \tag{S18}$$

Selecting $\frac{2\pi}{L_x}k_x^{(m)} = \frac{2\pi}{L_y}k_y^{(m)} = q_c$, we set all values of **q** in Eq.(18) within the square domain $-q_c \leq q_x, q_y \leq q_c$ and minimize the effect of non-square shape of the frame. Further averaging over orientation of **q** allows us to express $E_{el}$ through the elastic energy spectrum $E_{els}(q)$

$$E_{el} = \int E_{els}(q)dq \tag{S19}$$

$$E_{els}(q) = \frac{K}{2W(q)}\sum_{\mathbf{q'}} \tilde{Q}_{ij}(\mathbf{q'})\tilde{Q}_{ij}^*(\mathbf{q'})q'^2 w(q,q') \tag{S20}$$

is calculated as a sum over **q'** that are inside the ring of the width $2h$, $|\mathbf{q'} - \mathbf{q}| < h$, $w(q,q') = 1 - \frac{(q-q')^2}{h^2}$ is the weighting function, and $W(q) = 2\pi \int_{q-h}^{q+h} w(q,q')q'dq' = \frac{8}{3}\pi qh$. By introducing the weighted summation over $q$,

$$\sum_{\mathbf{q'}}^{w} = \frac{1}{W(q)}\sum_{\mathbf{q'}} w(q,q')$$

We obtain the equation for the elastic energy spectrum in the main text,

$$E_{els}(q) = \frac{K}{2}\sum_{\mathbf{q'}}^{w} \tilde{Q}_{ij}(\mathbf{q'})\tilde{Q}_{ij}^*(\mathbf{q'})q'^2,$$

In our case, $\bar{N} = 4081$, $L_x = 720\ \mu m$, $L_y = 480\ \mu m$, $k_x^{(m)} = 38$, $k_y^{(m)} = 26$. Thus, $\frac{2k_x^{(m)}+1}{2k_y^{(m)}+1} = 1.45 \approx \frac{L_x}{L_y} = 1.5$.

### 11.2. *Kinetic Energy Spectrum.*

The velocity of the $l$-th bacterium is defined by the vector $\mathbf{v}^{(l)} = \left(v_x^{(l)}, v_y^{(l)}\right)$, $l = 1, \ldots, N$. Its DFT reads

$$\tilde{v}_i(\mathbf{q}) = \frac{1}{N}\sum_{l=1}^{N} e^{-i\mathbf{q}\cdot\mathbf{r}_l}v_i^{(l)}, \mathbf{q} = \left(\frac{2\pi}{L_x}k_x, \frac{2\pi}{L_y}k_y\right), \tilde{v}_i(-\mathbf{q}) = \tilde{v}_i^*(\mathbf{q}), \tag{S21}$$

where $\mathbf{r}_l$ is the location of the $l$-th bacterium, $i = x, y$,

$$k_x = \left(-k_x^{(m)}, \ldots, k_x^{(m)}\right), \ k_y = \left(-k_y^{(m)}, \ldots, k_y^{(m)}\right), \bar{N} = \left(2k_x^{(m)} + 1\right)\left(2k_y^{(m)} + 1\right) \leq N, \tag{S22}$$

where $k_x^{(m)}$ and other variables are defined as above. The inverse DFT is defined as

$$\bar{v}_i^{(l)} = \sum_{k_x=-k_x^{(m)}}^{k_x^{(m)}} \sum_{k_y=-k_y^{(m)}}^{k_y^{(m)}} \tilde{v}_i(\mathbf{q})e^{i\mathbf{q}\cdot\mathbf{r}_l} = \sum_{\mathbf{q}} \tilde{v}_i(\mathbf{q})e^{i\mathbf{q}\cdot\mathbf{r}_l} \tag{S23}$$

Note that $\bar{v}_i^{(l)} = v_i^{(l)}$ if $\bar{N} = N$. Continuous extension $\mathbf{r}_l \to \mathbf{r}$ and $\bar{v}_i^{(l)} \to v_i(\mathbf{r})$ yields

$$v_i(\mathbf{r}) = \sum_q \tilde{v}_i(\mathbf{q})e^{i\mathbf{q}\cdot\mathbf{r}} = \sum_{k_x=-k_x^{(m)}}^{k_x^{(m)}} \sum_{k_y=-k_y^{(m)}}^{k_y^{(m)}} e^{i\mathbf{q}\cdot\mathbf{r}} \tilde{v}_i(\mathbf{q}). \tag{S24}$$

The kinetic energy spectrum $K_s(q)$ is defined through the area averaged velocity field $\langle \mathbf{v}^2 \rangle = \frac{1}{L_x L_y} \int \mathbf{v}^2 \, d\mathbf{r} = 2 \int K_s(q) dq$. Using Eq.(S24) and orthogonality of the Fourier harmonics, the area averaged velocity field $\langle \mathbf{v}^2 \rangle$ can be defined in the discrete form as

$$\langle \mathbf{v}^2 \rangle = \sum_q \left( \tilde{v}_x(\mathbf{q})\tilde{v}_x^*(\mathbf{q}) + \tilde{v}_y(\mathbf{q})\tilde{v}_y^*(\mathbf{q}) \right) = \sum_{k_x=-k_x^{(m)}}^{k_x^{(m)}} \sum_{k_y=-k_y^{(m)}}^{k_y^{(m)}} \left( \tilde{v}_x(\mathbf{q})\tilde{v}_x^*(\mathbf{q}) + \tilde{v}_y(\mathbf{q})\tilde{v}_y^*(\mathbf{q}) \right). \tag{S25}$$

Then, the kinetic energy spectrum $K_s(q)$ is calculated as

$$K_s(q) = \frac{1}{W(q)} \sum_{\mathbf{q}'} \left( \tilde{v}_x(\mathbf{q}')\tilde{v}_x^*(\mathbf{q}') + \tilde{v}_y(\mathbf{q}')\tilde{v}_y^*(\mathbf{q}') \right) w(q,q'). \tag{S26}$$

### 11.3. *Enstrophy Spectrum.*

The enstrophy area density is defined by the curl of bacterial velocities, integrated over the frame area:

$$\Omega = \frac{1}{L_x L_y} \int (\text{curl } \mathbf{v})^2 \, d\mathbf{r} \tag{S27}$$

Using Eq.(S25) and orthogonality of the Fourier harmonics, the enstrophy area density can be defined in the discrete form as

$$\Omega = \sum_q \left| \tilde{v}_x(\mathbf{q})q_y - \tilde{v}_y(\mathbf{q})q_x \right|^2 = \sum_{k_x=-k_x^{(m)}}^{k_x^{(m)}} \sum_{k_y=-k_y^{(m)}}^{k_y^{(m)}} \left| \tilde{v}_x(\mathbf{q})q_y - \tilde{v}_y(\mathbf{q})q_x \right|^2. \tag{S28}$$

Selecting $\frac{2\pi}{L_x} k_x^{(m)} = \frac{2\pi}{L_y} k_y^{(m)} = q_c$, we set all values of $\mathbf{q}$ in Eq.(28) within the square domain $-q_c \leq q_x, q_y \leq q_c$ and minimize the effect of the non-square shape of the frame. Averaging over the orientations of $\mathbf{q}$ allows us to express $\Omega$ through the enstrophy spectrum $\Omega_s(q)$:

$$\Omega = \int \Omega_s(q) dq, \tag{S29}$$

where

$$\Omega_s(q) = \frac{1}{W(q)} \sum_{\mathbf{q}'} \left| \tilde{v}_x(\mathbf{q}')q'_y - \tilde{v}_y(\mathbf{q}')q'_x \right|^2 w(q,q') = \sum_{\mathbf{q}'}^w \left| \tilde{v}_x(\mathbf{q}')q'_y - \tilde{v}_y(\mathbf{q}')q'_x \right|^2 \tag{S30}$$

is calculated as a sum over $\mathbf{q}'$ that are inside the ring of the width $2h$, $|\mathbf{q}' - \mathbf{q}| < h$, $w(q,q')$ and $W(q)$ is the normalization factor. In our case, $h = 3 \cdot \frac{2\pi}{L_x} = 0.0262 \ \mu m^{-1}$, $q \in \left[ \frac{2\pi}{240}, \frac{2\pi}{10} \right] \mu m^{-1}$, as $q_{min} \geq h$, $q_{max} \geq q_c - h$.

## *12. Number fluctuations of bacteria*

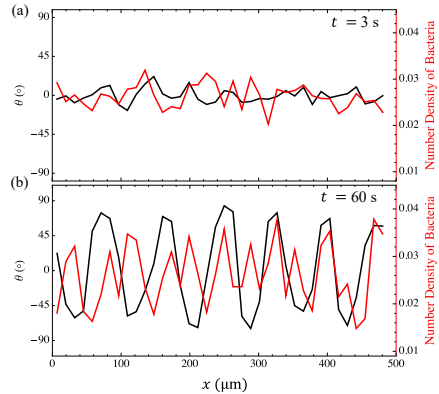

*Fig. S9. Number fluctuations of B. Subtilis in a sample with bacteria concentration* $5c_0$. *The average orientation angle θ and number density as a function of the x-axis at (a)* $t = 3$ *s and (b) 40 s.*